\documentclass[10pt]{article}
\usepackage[T1]{fontenc}
\usepackage{blindtext}
\usepackage{epigraph}
\usepackage{lipsum}
\usepackage{authblk}
\usepackage{lineno}
\usepackage[labelfont=bf]{caption}
\usepackage[numbers]{natbib}
\usepackage[resetlabels]{multibib}
\newcites{sup}{Supplementary References}
\usepackage{subcaption}
\usepackage[dvipsnames]{xcolor}
\usepackage{listings}
\usepackage{tcolorbox}
\tcbuselibrary{listings,breakable,skins}

\lstdefinestyle{prompttext}{%
  basicstyle=\fontfamily{lmtt}\fontsize{8.6}{10.8}\selectfont
             \color{black!85},
  columns=fullflexible,
  keepspaces=true,
  showstringspaces=false,
  breaklines=true,
  breakatwhitespace=true,
  breakindent=0pt,
  breakautoindent=false,
  aboveskip=0pt,
  belowskip=0pt
}

\newtcblisting{promptbox}[1]{%
  enhanced jigsaw,%
  breakable,%
  listing only,%
  listing engine=listings,%
  listing options={style=prompttext},%
  frame hidden,%
  boxrule=0pt,%
  arc=0pt,%
  colback=white,%
  colbacktitle=white,%
  coltitle=black!65,%
  fonttitle=\sffamily\bfseries\footnotesize,%
  titlerule=0pt,%
  borderline west={0.8pt}{0pt}{black!35},%
  left=3.5mm,%
  right=2mm,%
  top=2mm,%
  bottom=2mm,%
  left skip=\parindent,%
  right skip=\parindent,%
  before skip=11pt,%
  after skip=12pt,%
  title={#1}%
}

\usepackage{graphicx}
\usepackage{tabularx}
\usepackage{booktabs}
\usepackage{pdflscape}
\usepackage{adjustbox}
\usepackage{fvextra}
\usepackage{needspace}

\graphicspath{ {./figures/} }
\newcommand\blfootnote[1]{%
  \begingroup
  \renewcommand\thefootnote{}\footnote{#1}%
  \addtocounter{footnote}{-1}%
  \endgroup
}

\usepackage[colorlinks=true,linkcolor=Blue, urlcolor  = Blue, citecolor=Blue]{hyperref}%
\newcommand{\figref}[1]{Figure~\hyperref[#1]{\ref*{#1}}}
\newcommand{\figpanelref}[2]{Figure~\hyperref[#1]{\ref*{#1}#2}}
\newcommand{\figpanelrangeref}[3]{Figures~\hyperref[#1]{\ref*{#1}#2--#3}}
\newcommand{\figrangeref}[2]{Figures~\hyperref[#1]{\ref*{#1}}--\hyperref[#2]{\ref*{#2}}}
\usepackage{geometry}
\newcommand{\manuscriptwordcount}{%
  \IfFileExists{\jobname.wordcount.tex}{2961}{??}%
}
 \vspace{.5in}
\title{\LARGE \textbf{Gender and the Production of Research Impact
}\vspace{0.125in}}

\author[1,2]{Sander Wagner}
\author[1,2]{Charles Rahal}
\author[1,2,3]{Melinda C. Mills}

\affil[1]{Leverhulme Centre for Demographic Science, University of Oxford, United Kingdom}
\affil[2]{Nuffield College, University of Oxford, United Kingdom}
\affil[3]{Department of Econometrics, Economics and Finance, University of Groningen, Netherlands}

\begin{document}

\maketitle
\blfootnote{\noindent\textbf{For correspondence: }
\href{mailto:sander.wagner@demography.ox.ac.uk}{Sander Wagner}
or \href{mailto:charles.rahal@demography.ox.ac.uk}{Charles Rahal}. Address: Leverhulme Centre for Demographic Science, Demographic Science Unit, 42-43 Park End Street, Oxford, Oxfordshire, UK, OX29NR. Tel: 01865 743660. \textbf{Code Availability Statement:} All code used to generate the results in this paper is available in the public \texttt{ref\_gender} repository maintained by the Leverhulme Centre for Demographic Science (\url{https://github.com/OxfordDemSci/ref_gender}). \textbf{Data Availability Statement:} All REF2021 data are publicly available through the REF results database and are available under a Creative Commons Attribution 4.0 International (CC BY 4.0). The raw REF2021 data -- both structured and unstructured sources -- are also ingested on the fly by our code library, and merged with the Dimensions bibliometric database, provided upon licence for the purpose of analysing REF2021. \textbf{Acknowledgements:} Funding is gratefully acknowledged from the Leverhulme Trust (Grant RC-2018-003) for the Leverhulme Centre for Demographic Science, and Nuffield College. We are grateful to colleagues at the Nuffield College Feedback Working Group and the Social Sciences Impact Conference 2026 held at the University of Oxford for helpful comments on earlier versions of this manuscript.
}

\noindent\textbf{Abstract:} Evaluating the impact of scientific research beyond academia—on policy, health, the economy, and cultural life—has become a cornerstone of science policy and research-funding allocation worldwide. Yet which researchers produce the research underpinning this impact, and how this production is shaped by gender, remains poorly understood. We combine structured and unstructured records from the United Kingdom's latest Research Excellence Framework, the largest national research assessment currently in operation, with large-scale bibliometric data to quantify gender differences among the researchers underpinning documented impact. Women account for 38.16\% of these contributors: underrepresented overall, but  with a consistently higher share than in research-output authorships (33.63\%), both overall and across all four REF panels. Impact production is also strongly gendered across domains: women are better represented in case studies concerning education, health, cultural, and civil-society impact, whereas those concerning commercialisation pathways such as patenting and manufacturing remain dominated by men. These findings reveal critical inequalities across the pathways that connect research to impact beyond academia, offering crucial evidence for policymakers and academic institutions aiming to build more equitable and representative systems for evaluating scientific contributions. \vspace{.1in}

 \vspace{.1in}


\newpage

Who produces research that ultimately changes society? As governments and research funders increasingly demand evidence of societal and economic impact, understanding which researchers produce the research underpinning documented impact has become a central question for science policy. Despite decades of work documenting gender gaps in scientific careers, outputs, and funding acquisition \cite{Huang2020, Lariviere2013, witteman2019gender}, far less attention has been paid to the role of women in transforming scientific labour into tangible benefits beyond academia. Research has established that women remain underrepresented across many fields and career stages, particularly in senior positions and in Science, Technology, Engineering and Medicine (STEM) disciplines \cite{Holman2018}. They face disadvantages in pay and funding \cite{Shen2013}, citations \cite{lerman2022gendered}, authorship credit \cite{Ross2022}, and cumulative research productivity \cite{Huang2020}. Explanations for these gaps point to structural barriers that compound over time. Unequal family responsibilities contribute to shorter and more disrupted career trajectories \cite{Derrick2022}; differential attribution of credit generates disadvantages in recognition and advancement \cite{Ross2022}. Limited access to senior positions, in turn, constrains opportunities for sustained scientific output \cite{XieShauman1998}. Much of what is known about gender disparities in science is derived from analyses of established scholarly metrics, including publications \cite{Lariviere2013}, funding outcomes \cite{bedi2012gender}, and citation patterns \citep{lerman2022gendered}. These measures capture important dimensions of scientific productivity but fail to reflect the full range of contributions that researchers make. Governments and scientific funders increasingly value research impact \cite{Reed2021}: the demonstrable benefits of research for policy, health, the environment, the economy, cultural life, or quality of life more broadly. While qualitative evidence based on evaluator accounts suggests that the production of impact is itself highly gendered \cite{Chubb2020}, there remains remarkably little quantitative evidence on who produces the research underpinning documented impact, how this production is distributed by gender, and how it relates to more conventional measures of academic productivity.


\subsection*{A Window into Impact: Evidence from a national research evaluation}

The United Kingdom's Research Excellence Framework (REF) is the largest national research assessment system currently in operation, allocating approximately £2 billion annually across its seven-year assessment cycle on the basis of three evaluated components: research outputs, research environment, and impact beyond academia \cite{pinar_assessing_2022}. In REF of 2021 (REF2021), impact accounted for 25\% of the overall score for each submission. For the impact component, higher education institutions submit detailed Impact Case Studies (ICSs) that document how specific research generated demonstrable benefits beyond academia. Case studies are submitted to thematically organised Units of Assessment (UoAs), which are grouped into four broad disciplinary panels spanning medicine and life sciences (Panel A), physical sciences and engineering (Panel B), social sciences (Panel C), and arts and humanities (Panel D). Each UoA is assessed by a panel comprising academic reviewers and research users, who score ICSs according to standardised criteria of reach and significance. A total of 6,781 ICSs were submitted to REF2021 and a total of 6,361 ICSs from 157 institutions across 1,850 combinations of higher education institutions and UoAs were made publicly available \cite{ref2021_impact_database}. This makes it the largest coherent collection of structured impact narratives currently available internationally \cite{Wagner2024}. ICSs provide a rare and rich window into the generation of impact: they describe the underpinning research, identify contributing staff, and document the pathways through which research has influenced public policy, professional practice, industry, culture, health, the environment, or civil society (as described in Section~\ref{sec:si-impact-case-studies}). By combining the scale of a national evaluation exercise with detailed information on contributors and impact pathways, ICSs offer a distinctive empirical lens on the science--society interface. They enable systematic analysis of impact as a form of scientific labour and provide a foundation for assessing the gendered organisation of impact production across contemporary research systems. At the same time, they reflect submitted and evaluated instances of impact rather than the full universe of societal effects produced by research. While this introduces an element of selection, the near-complete coverage of UK higher education institutions and the breadth of submissions make REF the most comprehensive and systematically collected source currently available for studying societal impact at scale.

Impact case studies do not identify everyone who created or delivered impact: their staff fields identify researchers who conducted the underpinning research. We augment the ICSs to ask four questions: (i) how women's representation among these researchers varies across academic domains; (ii) whether it mirrors representation in research output (Section \ref{sec:si-research-outputs}); (iii) whether it varies across types of impact; and (iv) how academic domains, impact types, and institutional characteristics (Section \ref{sec:si-institutional-characteristics}) jointly shape it. We link the case studies to bibliometric records (Section~\ref{sec:si-dimensions}) and institutional data on UK higher education providers. Contributor names are recovered from structured output records and unstructured ICS PDFs using large language models and regular expressions (Section~\ref{sec:si-staff-parsing}), then matched to name-based gender inference (Section~\ref{sec:si-gender-inference}). We classify case studies into impact domains using large language models and validate them through human review of a sample, regular-expression rules, and alternative LLM classifications (Sections~\ref{sec:si-impact-domains}--\ref{sec:si-impact-domain-validation}). We analyse the resulting data descriptively and with regression models (Section~\ref{sec:si-domain-analyses}), supported by semi-structured interviews with REF panel members (N=36) and impact case study authors (N=12) conducted between March and December 2023 (Section~\ref{sec:si-qualitative-interviews}).

\subsection*{Absolute and Relative Representation of Women in the Production of Research Impact}

We identify 17,772 authorships naming underpinning researchers in 6,351 of the 6,361 published ICSs (99.84\%). Gender labels were inferred from given names, rather than self-reported, for 16,446 authorships (92.5\%); women account for 38.16\% of them, indicating clear under-representation overall.  As \figpanelref{fig:gender_ratios}{a} shows, however, this under-representation varies sharply across academic domains. In some UoAs -- such as Social Work and Social Policy (57.30\%), English Language and Literature (57.17\%), and Allied Health Professions, Dentistry, Nursing and Pharmacy (57.15\%) -- women constitute the majority of named underpinning-research contributors. In others -- notably Physics (14.05\%), Engineering (16.06\%), and Computer Science and Informatics (16.22\%) -- they remain a small minority. The lowest levels of women's representation are concentrated in the physical sciences and closely related fields, indicating that representation in the research underpinning documented impact reflects the broader gender structure of scientific employment. However, this representation is not gendered in exactly the same way as research output. Overall, women's representation is 4.52 percentage points higher in ICS authorships than in submitted research-output authorships (38.16\% versus 33.63\%). This pattern is visible at the level of the four REF main panels in \figpanelref{fig:gender_ratios}{b}; women are more strongly represented in ICS authorships than in outputs in all four panels, with the physical sciences (Panel~B) standing out for its low representation of women on both measures. Impact production therefore appears comparatively more gender-balanced than research output, even though women remain under-represented overall. Comparisons at the level of individual UoAs further underscore the unevenness of this pattern. \figpanelref{fig:gender_ratios}{d} shows the ratio of women's representation in ICS authorships relative to outputs across fields. In most UoAs, women are more strongly represented among named underpinning-research contributors than among output authors, with 21 out of 34 fields situated above parity. Importantly, heavily male-dominated fields appear on both sides of this divide: Philosophy and Physics show more representation in ICS authorships relative to outputs, whereas Engineering and Chemistry show the opposite pattern. This divergence suggests that overall field composition alone cannot explain gender differences in the research underpinning documented impact. Instead, it points to systematic differences in the kinds of impact produced across fields. Supplementary descriptives are reported in Section~\ref{sec:si-descriptive-results}. Here and throughout, ICS authorships are contributor mentions on the case studies; they identify the staff credited with conducting the underpinning research, not every person involved in creating or delivering the eventual impact.

\subsection*{Impact is not a Single Activity: Strong gender sorting across impact domains}
 
The divergence observed across academic fields highlights a central feature of research impact that is obscured in aggregate analyses: impact is not a single activity, but comprises a heterogeneous set of practices, objectives, and domains. ICSs document engagement with a wide range of non-academic actors and outcomes, including education and public outreach, public policy and regulation, health and clinical practice, cultural production, civil society, and commercial or industrial applications. The relative prominence of these domains varies systematically across fields. For example, impact in philosophy and physics more often centres on education and scientific outreach, whereas chemistry and engineering more frequently emphasise industrial applications and patenting. These differences provide a plausible explanation for why male-dominated fields can exhibit sharply different gender patterns in impact relative to outputs. To examine these patterns systematically, we identify a set of recurrent impact domains and assign individual ICSs to these domains using large language models (Section~\ref{sec:si-impact-domains}; \figref{fig:supp_topic_validation} compares assignments across large language models and regular-expression rules). \figpanelref{fig:fig2}{a} shows the share of women among named underpinning-research contributors within each domain, disaggregated by REF main panel. 
Clear gender sorting is evident across impact domains. Women are disproportionately represented in impact activities centred on education, culture, public engagement, and work with public-sector and civil-society organisations. These forms of impact typically involve knowledge dissemination, advisory roles, and sustained interaction with non-academic audiences. By contrast, men are more strongly represented in impact linked to industrial collaboration, manufacturing, commercialisation, and patenting, which tend to be embedded in private-sector partnerships and market-oriented outcomes. A complementary text-based analysis, which asks only whether individual words in ICS texts co-occur with higher or lower shares of women authors, without imposing any predefined domain classification, recovers the same pattern: terms associated with commercialisation and industry co-occur with lower shares of women authors, while terms related to education, community engagement, and social outcomes co-occur with higher shares (\figref{fig:supp_text}). The convergence between our domain classification and this fully data-driven textual analysis supports our finding of highly gendered impact domains.

\begin{figure}[t!]
\centering
\includegraphics[width=\textwidth]{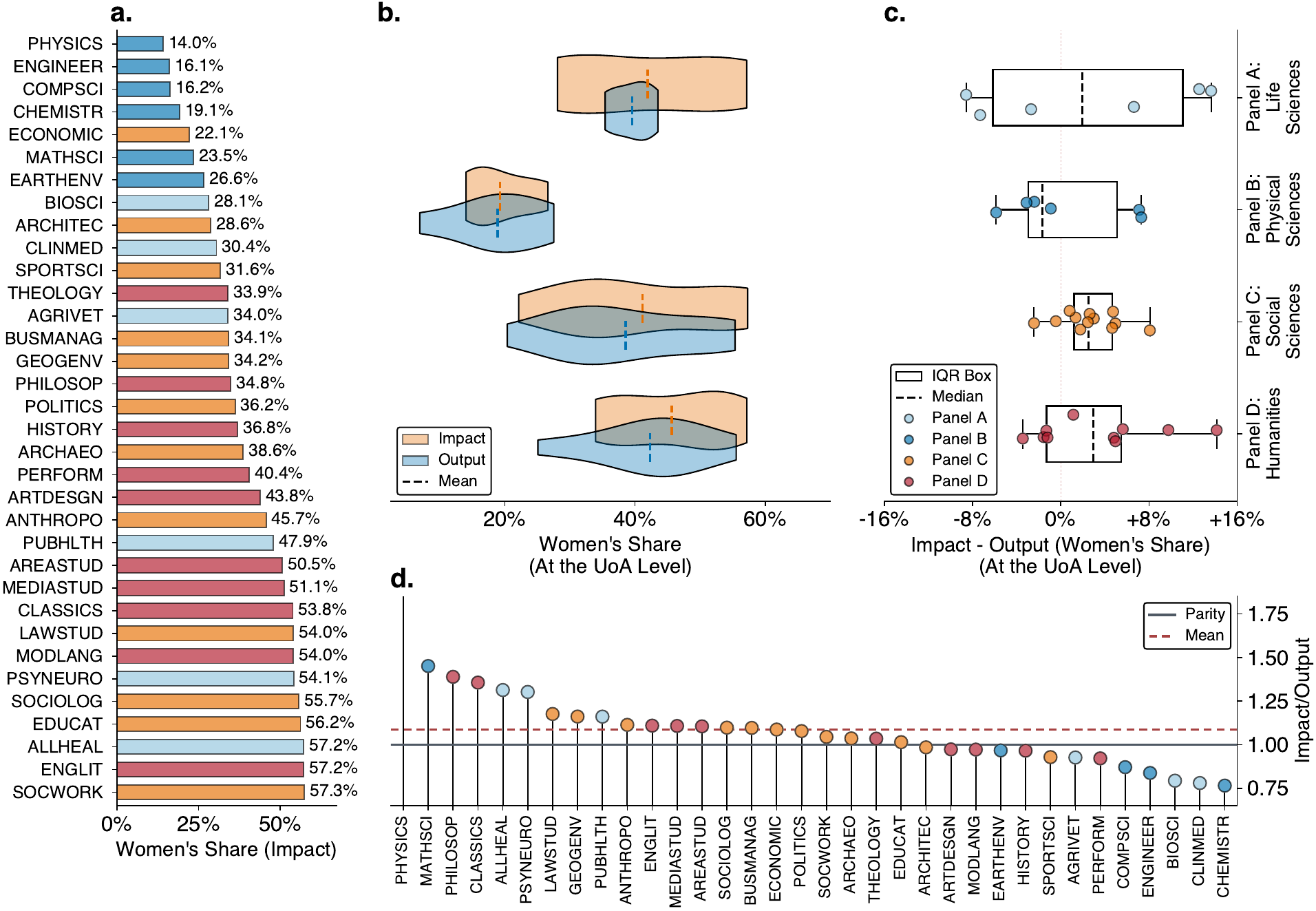}
\caption{
\textbf{Gender representation among named underpinning researchers in impact case studies and among research-output authors across REF2021 fields.} Subfigure `\textbf{a.}' shows women's share among researchers named as conducting the underpinning research in ICSs by UoA. Subfigure `\textbf{b.}' shows the UoA-level distributions of women's shares among named ICS underpinning researchers and research-output authors across the four REF main panels. Subfigure `\textbf{c.}' shows the difference between these two measures across UoAs. Subfigure `\textbf{d.}' shows their ratio; values above one indicate greater women's representation among named ICS underpinning researchers than among output authors. IQR represents the Inter-Quartile Range. Abbreviations and figure labels are described in Section~\ref{sec:si-abbreviations} and Tables~\ref{tab:supp-abbreviations}-\ref{tab:supp-uoa-labels}.
}
\label{fig:gender_ratios}
\end{figure}

The magnitudes are substantial. Women account for nearly half of authors in Charity (47.54\%), School (47.04\%), Museum (46.49\%), and NHS (45.45\%) case studies, but only around one in five in Patent (17.89\%), Manufacturing (20.02\%), and Startup (22.83\%) domains. These are not marginal categories: Charity, Software, Legislation, School, and NHS are the five most frequently assigned impact domains. 
Overall and panel-specific domain descriptives are reported in Tables~\ref{tab:supp-llm-summary} and \ref{tab:supp-llm-panel-summary}. A focused comparison of Physics and Chemistry illustrates the pattern concretely. Both fields have very low representation of women in research outputs and in impact case studies overall (Table~\ref{tab:supp-uoa-summary}), yet Physics shows relatively higher women's representation in impact compared with outputs than Chemistry does (\figpanelref{fig:gender_ratios}{d}). This divergence is consistent with differences in domain composition: Physics has a higher prevalence of School and Museum case studies (domains positively associated with women's representation), whereas Chemistry concentrates in Manufacturing, Patent, and Drug Trial (negatively associated); the relevant Model Three coefficients are reported in Table~\ref{tab:regressions}. Our regression-based approach (Section~\ref{sec:si-domain-analyses} and \figpanelref{fig:fig2}{b}) indicates a domain-predicted component of \(-2.9\) percentage points for Physics, compared with \(-8.5\) percentage points for Chemistry (Table~\ref{tab:physics_chemistry_domains}), consistent with part of the observed difference between the two fields. At the same time, \figpanelref{fig:fig2}{a} shows that domain composition alone does not fully account for gender differences among named underpinning-research contributors. Within every impact domain, the share of women among these contributors is systematically lower when case studies originate from the STEM-intensive Panel~B. This pattern indicates that differences associated with disciplinary context remain on top of those associated with the type of impact pursued. Understanding gender differences in impact production therefore requires accounting jointly for impact domains, academic fields, and institutional context.

\subsection*{Domains, Disciplines, and Institutions in Impact Production}
 
\begin{figure}[t!]
\centering
\includegraphics[width=\textwidth]{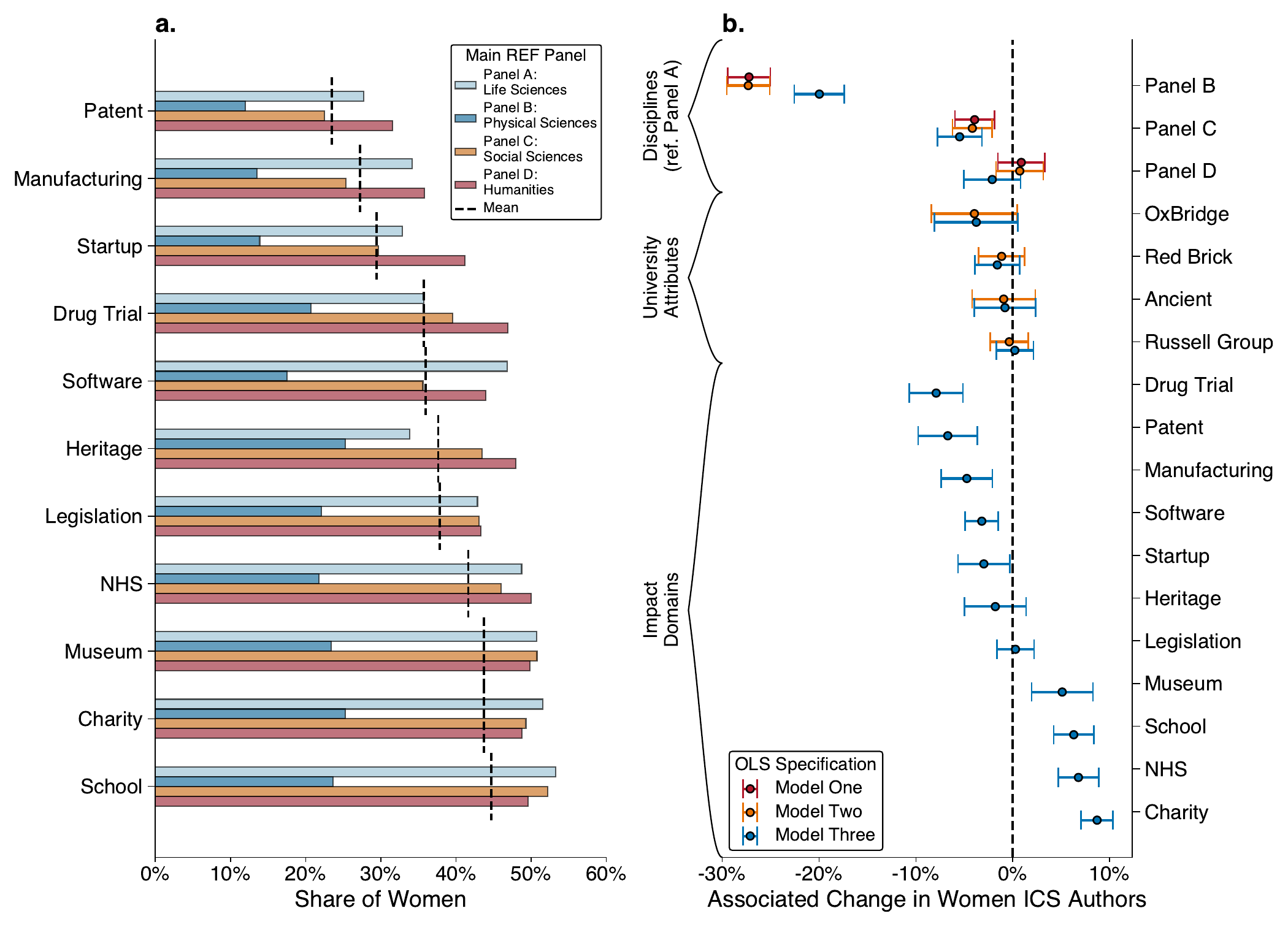}
\caption{
\textbf{Gender representation among named underpinning researchers across impact domains.} Subfigure `\textbf{a.}' shows women's share among researchers named as conducting the underpinning research in ICSs by impact domain, disaggregated by REF main panel (Panel A: Life Sciences; Panel B: Physical Sciences and Engineering; Panel C: Social Sciences; Panel D: Arts and Humanities). Domains are ordered from lowest to highest overall women's share. The vertical dashed black line indicates the mean women's share across all case studies. \figref{fig:supp_domain_models} varies this across classification strategy. Subfigure \textbf{`b.'} shows coefficients from OLS regression models of women's share among these named researchers, estimated across three nested specifications. Model One includes disciplinary panel indicators only (reference category: Panel A). Model Two adds university-type indicators (Oxbridge, Russell Group, Red Brick, Ancient). Model Three further adds binary indicators for each impact domain (reference category: no domain assigned). Markers show point estimates; horizontal bars show 95\% confidence intervals. All models are estimated on 6,173 ICSs with at least one named researcher of identifiable gender; Table~\ref{tab:regressions} shows full regression results for all three models when estimated with both ordinary least squares and generalised linear models.
}
\label{fig:fig2}
\end{figure}

\figpanelref{fig:fig2}{b} summarises results from three different regression specifications that incrementally control for disciplinary context (Model One), institutional characteristics (Model Two), and impact domains (Model Three). The largest estimated associations are with disciplinary composition and the gendered organisation of impact domains, while those with institutional characteristics are comparatively small. Regression results remain highly stable across model specifications, estimators, and ways of measuring disciplinary composition (\figrangeref{fig:supp_glm}{fig:supp_uoa_models}; Table~\ref{tab:regressions}). Disciplinary context remains a strong predictor of women's participation in impact. When impact domains are not taken into account, case studies from the STEM-intensive Panel~B are associated with a \(27.2\) percentage-point lower share of women authors (95\% CI, \(-29.5\) to \(-25.0\)). Accounting for differences in impact domains attenuates this gap to \(20.0\) percentage points (95\% CI, \(-22.6\) to \(-17.4\)), but does not eliminate the strong conditional association between Panel~B and gender composition. Panel~B's gender imbalance therefore reflects both the types of impact pursued and discipline-specific norms and career structures that operate independently of impact content. Impact domains themselves show additional and independent associations: even after accounting for disciplinary composition, impact linked to startups, drug trials, and patenting is associated with substantially lower women’s representation, whereas impact related to museums, the NHS, and schools is associated with higher women's involvement.

\subsection*{Implications for Understanding Gender and Research Impact in Science}
 
Our analysis of published Impact Case Studies and research outputs submitted to REF2021 yields three interrelated findings. First, women remain underrepresented among the researchers whose work underpins documented impact. This underrepresentation broadly mirrors persistent gender inequalities across the UK research system and across scientific disciplines, particularly in STEM fields. Impact production, like research-output authorship, is embedded in existing structures of gender inequality rather than operating independently of them. Second, despite this overall underrepresentation, women are more strongly represented among ICS underpinning-research authorships than among research-output authorships overall and in each of the four REF panels. At the UoA level, this pattern holds in 21 of 34 fields. This pattern indicates that women's scientific contributions are relatively more concentrated in activities that generate verifiable benefits beyond academia than in those most strongly rewarded through publication-based metrics. Third, the production of research underpinning impact is itself highly gendered across both scientific disciplines and impact domains. STEM fields --- particularly those grouped in Panel~B --- display the lowest representation of women in both output and impact, yet also exhibit some of the most pronounced differences between the two. At the same time, impact pathways vary sharply in their gender composition. Case studies involving education, culture, the NHS, charities, and public services involve comparatively high shares of women, whereas those involving commercialisation-oriented impact --- such as manufacturing partnerships, industrial applications, and patenting --- overwhelmingly feature men. These patterns reveal that gender inequality in impact production reflects not only where women and men are positioned within the research system, but also how different forms of engagement and knowledge translation are organised and valued. The estimates describe mentions of researchers credited with conducting the underpinning research in the impact case studies, rather than every individual who created or delivered the impact, and are conditional on successful output linkage. Gender is inferred from names rather than self-reported, so these categories cannot capture how researchers identify, including non-binary identities.

Taken together, these findings have important implications for science policy and research evaluation. By shifting attention from publication metrics to research impact, evaluation systems bring into view dimensions of scientific contribution that are increasingly valued by funders and rarely analysed at scale. The strong gender sorting across impact domains indicates that impact is not a uniform corrective to existing inequalities, but a differentiated set of activities that may reproduce hierarchies in new ways. Women fare better in impact than on conventional output measures, yet remain concentrated in its least commercial pathways. Qualitative interviews with researchers and research managers point to several mechanisms that may help explain these patterns. Some interviewees noted that impact-related work can be more compatible with career interruptions linked to childbearing or caregiving, and that returning researchers, generally women, may find it easier to re-establish visibility through engagement and impact than through publication alone. This is visible in the very pathways where women are best represented. The experience of Jo Appleby — whose skeletal analysis underpinned the identification of King Richard III in one of the most prominent heritage case studies in REF2021 — illustrates both the character of impact work in these domains and its limits. In heritage, impact runs through exhibitions, public talks, and schools engagement, and she sustained an intensive programme of it until eight months pregnant with her first child: ``I was literally doing two after-work talks a week, and I could have done more [\ldots] it's kind of more than you should do when you're heavily pregnant, I realised afterwards.'' Yet two year-long periods of maternity leave still disrupted this work: ``losing that continuity caused some difficulty''.\footnote{Quoted by name with written permission; see Section~\ref{sec:si-qualitative-interviews}.} Impact may thus be more accommodating of career interruptions than publication-based output, but it is not exempt from them. Others described impact as offering alternative routes to recognition in highly gendered STEM environments where traditional prestige hierarchies remain difficult to navigate. While these accounts do not establish causal explanations, they underscore that impact work is embedded in career structures and institutional practices that shape who undertakes it and how it is rewarded.

\subsection*{Impact as Opportunity and Constraint}
 
As research impact becomes an increasingly central component of research assessment and funding, understanding its gendered organisation is critical. Our analysis shows that impact can function both as an opportunity and as a constraint in women's scientific careers. Women are relatively more involved in impact production than in research outputs, suggesting that impact offers pathways for engagement and recognition that extend beyond publication-based metrics. At the same time, women remain under-represented overall and are particularly excluded from the commercial and industrial pathways through which research connects to the economy. Whether impact contributes to narrowing or widening gender gaps in science therefore depends less on its formal inclusion in evaluation systems than on how different forms of impact and contributors to impact are valued, supported, and rewarded. Treating impact as a single category risks obscuring the gendered labour that underpins distinct pathways of economic and societal engagement. Designing evaluation systems that genuinely advance equity requires recognising impact as a heterogeneous form of scientific work and ensuring that the full range of contributions through which research benefits society is visible and valued. Practical first steps are for UK higher education funding bodies to require structured contributor and role fields in REF impact case studies, and for institutions to use these data to monitor gender representation across impact pathways and to address those, such as commercialisation, that remain disproportionately closed to women.

\bibliography{impact}

\clearpage
\newpage

\section*{Online Supplementary Materials}
\setcounter{subsection}{0}
\setcounter{subsubsection}{0}
\renewcommand{\thesubsection}{S\arabic{subsection}}
\renewcommand{\thesubsubsection}{\thesubsection.\arabic{subsubsection}}
\setcounter{table}{0}
\renewcommand{\thetable}{S\arabic{table}}
\setcounter{figure}{0}
\renewcommand{\thefigure}{S\arabic{figure}}

\subsection{Data Sources}\label{sec:si-data}

\subsubsection{Impact case studies}\label{sec:si-impact-case-studies}

We use the public REF2021 Impact Case Study (ICS) database, which contains 6,361 case-study records, as the source for analysing gender representation among researchers credited with underpinning documented impact \citesup{ref2021_impact_database_sup,ref2021_impact_database_faq}. We draw on it in two ways. First, for the case-study level analyses, we download the REF ICS export and retain the case-study identifier, institution, Unit of Assessment (UoA), REF main panel, title, and the five structured ICS text fields: summary of the impact, underpinning research, references to the research, details of the impact, and sources to corroborate the impact. These five text fields are the source material for the impact-domain indicators and the supplementary text analysis. Second, for the contributor-level gender analysis, we download and cache the corresponding public case-study PDFs, extract the block naming the staff who conducted the underpinning research (or an equivalent author block), and record case-level counts of women, men, unknown, and total extracted people. Cases that cannot be resolved are kept as unresolved in the staff-extraction audit rather than silently treated as
zero-person case studies.

\subsubsection{Research outputs}\label{sec:si-research-outputs}

We use the published REF2021 research-output workbook as the source list for the publication-output benchmark \citesup{ref2021_outputs_export}, downloading the workbook, reading it after the REF header rows, and retaining the REF output identifier, institution, UoA, REF main panel, DOI, and ISBN. Because the public REF output file does not provide a consistent author list suitable for author-level gender analysis, we enrich the REF output list with author metadata from Dimensions (Section~\ref{sec:si-dimensions}), using the REF-supplied DOIs and ISBNs to link the two sources. Specifically, we query Dimensions publications in DOI and ISBN batches, cache the returned metadata, normalise the DOI and ISBN strings, and merge the cached Dimensions records back to the REF output list, preferring the DOI match when both a DOI and an ISBN match are available. For matched outputs, we extract author first names from Dimensions and apply the same name-based gender inference procedure used for ICS contributors (Section~\ref{sec:si-gender-inference}), counting women, men, unknown, and total recovered authors for each output. This yields two tables: one containing all REF output rows, and one restricted to outputs with at least one recovered author. The main impact--output comparisons use the latter, aggregating the output-level counts to UoAs and REF main panels.

\subsubsection{Institutional characteristics}\label{sec:si-institutional-characteristics}

We use institutional information in two ways: as descriptive scale measures in the supplementary summary tables, and as institution-type indicators in the regression models. First, for the descriptive measures, we read the REF results workbook \citesup{ref2021_profiles_export}, standardise institution identifiers using UKPRN codes, construct institution--UoA identifiers, and retain the FTE of submitted staff. The UoA and REF main-panel summary tables aggregate these submitted-staff measures once per REF submission rather than once per impact case study. The descriptive tables also use two variables from the REF environment workbook \citesup{ref2021_environment_export}: doctoral volume, constructed from the \texttt{ResearchDoctoralDegreesAwarded} sheet by summing the annual columns recording doctoral degrees awarded, and research income, taken from the \texttt{ResearchIncome} sheet using the row labelled ``Total income''. These variables are reported in the summary tables at the same submission level as PhDs and total income. Second, for the regression models, we merge institution-type indicators from a manually curated lookup file in the analysis repository, identifying Oxbridge institutions, Russell Group membership, Red Brick universities, and Ancient universities; these four indicators serve as the institutional controls in the reported models. Red Brick universities are defined as the six large civic institutions founded in industrial cities in the late nineteenth and early twentieth centuries (Birmingham, Bristol, Leeds, Liverpool, Manchester, and Sheffield), and Ancient universities as institutions founded prior to 1600 (Oxford, Cambridge, St Andrews, Glasgow, Aberdeen, and Edinburgh). The categories are not mutually exclusive: Oxford and Cambridge are simultaneously Oxbridge, Ancient, and Russell Group institutions, while all Red Brick universities are members of the Russell Group.

\subsubsection{Dimensions from Digital Science}\label{sec:si-dimensions}

Dimensions is a research-information database from Digital Science that provides publication records, persistent identifiers, and author metadata  \citesup{hook_dimensions_2018,digital_science_dimensions}. In this study, we use it solely as a bibliographic enrichment source for the REF2021 research outputs: the REF output workbook supplies the list of submitted outputs together with, where available, their DOI and ISBN identifiers, and Dimensions supplies the corresponding author metadata, from which we recover the author first names used in the name-based gender inference procedure (Section~\ref{sec:si-gender-inference}). Concretely, we query the Dimensions Publications API through \texttt{dimcli} (an open source Python client for accessing the Dimensions Analytics API), separately in DOI and ISBN batches, requesting publication identifiers, DOI, ISBN, author lists, and author counts; returned records are cached locally so that downstream analyses can be rerun without repeating API calls. The normalisation of identifier strings and the merge back to the REF output list follow the procedure described in Section~\ref{sec:si-research-outputs}; where both a DOI and an ISBN match are available, the DOI match is preferred because DOIs are generally more specific to individual publication records.

\subsection{Data Extraction and Variable Construction}\label{sec:si-data-extraction}

\subsubsection{Parsing staff names from impact case studies}\label{sec:si-staff-parsing}
We extract contributor names for each impact case study from the public REF2021 case-study PDFs using the reproducible staff-extraction stage in the analysis pipeline. For each REF impact case study identifier, the pipeline locates the corresponding public case-study record, downloads the PDF when it is not already cached, and stores the cached copy for reproducible re-use. Text is extracted from each PDF using two parsers, PyMuPDF and \texttt{pdfminer}; when both parsers return text, the pipeline keeps the version that most clearly contains a staff-name block and, if neither parser exposes such a block, the longer usable text. This produces a case-level text file with the REF case-study identifier, extracted PDF text, the extracted staff block where found, and an extraction-status flag. The next stage isolates the part of the case-study document that identifies the people who conducted the underpinning research. The parser searches for the canonical REF ``Details of staff conducting the underpinning research'' section and for common variants of the \texttt{Name(s)}, \texttt{Author(s)}, \texttt{Role(s)}, and employment-period headers. Extraction stops at the next REF section marker, such as the start of the claimed-impact period, impact summary, underpinning-research section, impact-details section, or corroborating-sources section. Header variants are canonicalised to stable labels before downstream parsing. If the staff block is not found in the PDF text, the pipeline constructs a fallback source from the case-study title and the five structured REF text fields described in Section~\ref{sec:si-impact-case-studies}; this fallback is used only to recover named researchers associated with the underpinning research, not cited-paper authors, funders, beneficiaries, external partners, or people mentioned solely as impact users. Staff names are then parsed from the isolated block in two stages. First, a deterministic local parser reads explicit \texttt{Name(s)} or \texttt{Author(s)} sections, removes bracketed short descriptors, splits multiple names separated by line breaks, semicolons, and simple conjunctions, removes duplicates, and rejects non-person header or role fragments. If this local pass does not recover names and LLM extraction is enabled, the unresolved block is sent to the configured OpenAI staff-extraction model in the repository pipeline (in this version of the manuscript: \texttt{gpt-5.5}). Under the primary configuration the staff-extraction request is synchronous and single-case, so the exact system message sent to the model is:

\begin{promptbox}{Staff extraction: system message}
You are extracting named researchers/authors from REF impact case-study source text. Inputs may be a canonical 'Details of staff' table, an Author(s) block, or a full case-study text fallback. Prefer explicit Name(s)/Author(s) entries. For full text fallbacks, extract only researchers who conducted the underpinning research for the case study. Do not extract cited-paper authors, beneficiaries, funders, external partners, or people mentioned only as impact users. Return JSON {'people': [{'name': ..., 'roles': [...]}]}.
\end{promptbox}

\noindent The corresponding user message is the isolated \texttt{staff\_block}. When the PDF parser cannot isolate a staff block, the user message is constructed from the following exact prefix and then the case-study title and available structured REF text fields:

\begin{promptbox}{Staff extraction: fallback user-message template}
Fallback source: full REF impact case-study text. Extract the named researchers/authors who conducted the underpinning research for this case study. Do not extract cited-paper authors, beneficiaries, funders, external partners, or people mentioned only as impact users.

Title: <case-study title>

<structured REF case-study field name>
<structured REF case-study field text>
\end{promptbox}

The request also attaches the \texttt{emit\_staff} function tool, which requires a JSON object with a \texttt{people} array, each element containing a \texttt{name} string and optional \texttt{roles} strings. The extraction code supports batched requests, single-case fallback for missed batch items, and retries for transient API failures, but these are robustness paths rather than the prompt shape used in the primary staff-extraction configuration. A final deterministic local fallback is attempted after an empty or failed LLM response. For each recovered person, the pipeline normalises the name before gender inference. It strips common titles and post-nominals such as \texttt{Prof}, \texttt{Dr}, \texttt{Sir}, \texttt{Dame}, \texttt{PhD}, \texttt{DPhil}, \texttt{FRS}, and related honours; normalises case; and extracts a candidate given name by selecting the first token that is not an initial and not a common name particle such as \texttt{van}, \texttt{von}, \texttt{de}, \texttt{del}, \texttt{du}, or \texttt{da}. Formally, for a normalised name token sequence \(x=(t_1,\ldots,t_k)\), the candidate given name is
\[
  f(x)=\min\nolimits_i\{\,t_i:\ t_i\notin\mathcal I,\ t_i\notin\mathcal P\,\},
\]
where \(\mathcal I\) is the set of one- or two-letter initial patterns and \(\mathcal P\) is the particle set above; if no such token exists, the pipeline retains no given-name candidate for that person. The staff-extraction stage writes PDF-text, person-level, case-level, and audit tables. These outputs record the extracted text and staff blocks, one row per recovered person, case-level lists and counts, extraction status, extraction errors, and whether each case remains unresolved. Cases that remain genuinely unresolved are retained as unresolved with blank counts rather than converted to zero-person case studies.

\subsubsection{Name-based gender inference}\label{sec:si-gender-inference}
We infer gender labels, rather than observe self-reported gender identities, using two offline lexicons applied in fixed precedence order, both for the authors in our scientometric corpus of research outputs and for the researchers parsed from the impact case study staff blocks. For each person \(u\), let \(g_{gg}(u)\) denote the label returned by \texttt{gender\_guesser} and \(g_{gd}(u)\) the label returned by \texttt{gender\_detector[UK]}, each applied to the candidate given name constructed in Section~\ref{sec:si-staff-parsing}. Names consisting solely of initials or containing unresolved particles yield no candidate given name and are conservatively labelled \texttt{unknown}. For all other names, the final label \(G(u)\) is
assigned as follows:
\[
\begin{array}{l}
\textbf{if } g_{gg}(u)\neq\texttt{unknown} \textbf{ then}\\[4pt]
\quad G(u)\leftarrow g_{gg}(u)\\[6pt]
\textbf{else if } g_{gd}(u)\neq\texttt{unknown} \textbf{ then}\\[4pt]
\quad G(u)\leftarrow g_{gd}(u)\\[6pt]
\textbf{else}\\[4pt]
\quad G(u)\leftarrow \texttt{unknown}\\[4pt]
\textbf{end if}
\end{array}
\]
Both lexicons return richer label sets than we require, so we collapse them to three categories. For \texttt{\nolinkurl{gender_guesser}}, labels corresponding to men or mostly men are coded as \texttt{men}, labels corresponding to women or mostly women as \texttt{women}, and ambiguous or unknown labels as \texttt{unknown}. For  \texttt{\nolinkurl{gender_detector[UK]}}, labels corresponding to men and women are coded analogously, and all other or missing labels as \texttt{unknown}. We then re-aggregate the person-level labels to the case level by counting \texttt{men}, \texttt{women}, and \texttt{unknown} labels, and further to the institution and Unit of Assessment hierarchies using the raw REF Impact Case Study Data; retaining the ICS and paper identifiers throughout allows the data to be re-linked at any level. Applying this procedure to the 17,772 named ICS authorships yields a gender label for 16,446 (92.54\%), with 1,326 authorships (7.46\%) remaining \texttt{unknown}, including authorships represented only by initials. These categories should therefore be interpreted as name-based classifications, not observations of individual gender identity.

\subsubsection{Identification of impact domains using LLMs and regular expressions}\label{sec:si-impact-domains}

Impact-domain indicators are constructed as multi-label classifications of each REF impact case study. We first build a normalised text string from the five structured REF case-study fields: summary of the impact, underpinning research, references to the research, details of the impact, and sources to corroborate the impact. Each field is normalised by replacing dash variants, compacting whitespace, and retaining the field label before concatenation, so the model sees both the case-study text and the section from which each text segment originates. The eleven target domains are Charity, Startup, Patent, Museum, NHS, Drug Trial, School, Legislation, Heritage, Manufacturing, and Software. The same normalised text is classified in two ways: by a large language model, which provides the primary domain variables used in the analysis, and by deterministic regular-expression rules, which are retained as an independent rule-based comparison and as a fallback. The regular-expression indicators use case-insensitive keyword rules for each domain, with alphabetic word-boundary guards and allowance for common spacing or hyphenation variants. For example, the charity indicator matches terms for charities, non-profit organisations, NGOs, voluntary organisations, third-sector organisations, charitable trusts and foundations, and social enterprises, while the startup indicator matches startup, start-up, spinout, spin-out, spinoff, and spin-off variants. The LLM classification, in its current public implementation, uses \texttt{gpt-5.5}, the Responses API, service tier \texttt{flex}, one case study per request, a strict JSON schema requiring one result object per input identifier, and a deterministic cache key formed from the prompt version, model, and normalised case-study text. The model output is parsed into binary indicators, stored as integer columns, and reused from cache on subsequent runs. Where an LLM cache row is unavailable or invalid --- because the request failed, LLM extraction was disabled, or the response could not be parsed --- the corresponding regular-expression indicators are used instead, so that downstream code remains deterministic. The exact prompt template constructed by the implemented code is:

\begin{promptbox}{Impact-domain classification: prompt template}
[PROMPT_VERSION=v2]
You are a conservative multi-label classifier for REF impact case studies.
Given ITEMS (a JSON array with fields `id` and `text`), classify each item.

Decision rules:
1) Mark true only if that theme is materially involved in the impact claim.
2) Passing mention, background context, or weak association => false.
3) Multiple true labels are allowed.
4) If uncertain, choose false.

Indicator definitions:
  - charity: True only when charities/NGOs/third-sector actors are a material impact route, partner, or beneficiary.
  - startup: True only when startup/spinout creation, growth, or deployment is part of the impact pathway.
  - patent: True only when patents/patenting/licensing are materially involved in the impact.
  - museum: True only when museums/galleries/exhibitions are direct impact venues, partners, or beneficiaries.
  - nhs: True only when NHS bodies, services, pathways, or policy/practice are materially affected.
  - drug_trial: True only when drug/therapeutic development or clinical trial activity is materially involved.
  - school: True only when school-level policy/practice/curriculum/outcomes are materially affected.
  - legislation: True only when law/regulation/statutory guidance is created, changed, or implemented as impact.
  - heritage: True only when heritage institutions, assets, conservation policy, or practice are materially affected.
  - manufacturing: True only when industrial manufacturing processes, plants, or production outcomes are materially affected.
  - software: True only when software tools/platforms/systems are central to the delivered impact.

Return only valid JSON in this structure:
{
  "results": [
    {
      "id": "<id-from-input>",
      "charity": true/false,
      "startup": true/false,
      "patent": true/false,
      "museum": true/false,
      "nhs": true/false,
      "drug_trial": true/false,
      "school": true/false,
      "legislation": true/false,
      "heritage": true/false,
      "manufacturing": true/false,
      "software": true/false
    }
  ]
}
Include exactly one result object per input item id.

ITEMS:
[{"id": "<cache-key>", "text": "<normalised REF case-study text>"}]
\end{promptbox}

\noindent Only the final JSON array following \texttt{ITEMS} varies across requests: under the current configuration it contains one object whose \texttt{id} is the deterministic cache identifier and whose \texttt{text} is the normalised five-field case-study text. The prompt is deliberately conservative: labels are positive only when the domain is materially involved in the claimed impact, passing mentions are coded false, multiple domains can be true for the same case study, and uncertainty is resolved as false.

\subsubsection{Validation and cross-checks for impact-domain assignment}\label{sec:si-impact-domain-validation}
\begin{figure}[!ht]\centering
\includegraphics[width=\textwidth]{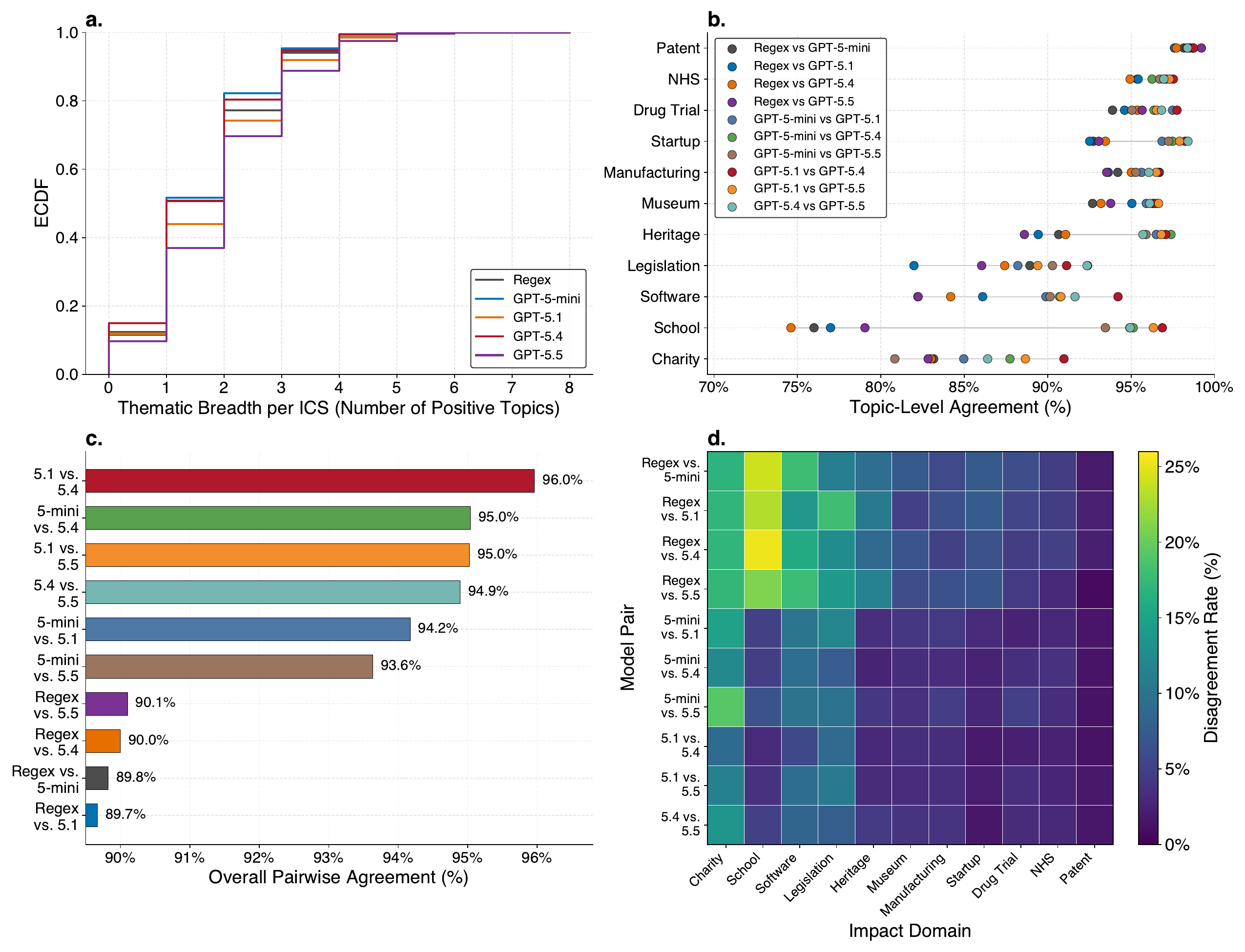}
\caption{
\textbf{Comparison of impact-domain classification across different LLMs and regular expression methods}. Subfigure `\textbf{a.}' shows the empirical cumulative distribution of the number of impact domains assigned per case study (thematic breadth), highlighting similar distributions across regular expression and LLM-based approaches. Subfigure `\textbf{b.}' shows topic-level agreement across classification methods, indicating consistently high agreement rates. Subfigure `\textbf{c.}' aggregates this across domains to highlight individual approach comparisons. Subfigure `\textbf{d.}' shows disagreement rates by domain and method pair, but with a different aesthetic approach.
}
\label{fig:supp_topic_validation}
\end{figure}

To assess the robustness of the impact-domain classification, we manually reviewed a sample of GPT-5.5 assignments against the underlying case-study text and compared the full set of assignments with both rule-based regular-expression classifications and those produced by alternative LLM variants; \figref{fig:supp_topic_validation} reports these checks. Subfigure `\textbf{a.}' shows that the number of domains assigned to each case study is very similar across approaches, indicating that the methods recover comparable levels of thematic breadth rather than systematically over- or under-classifying cases. Subfigures `\textbf{b.}'--`\textbf{d.}' show that the individual domain allocations are also stable: across all case-study--domain decisions, the regular-expression classifications agree with the GPT-5.5 classifications in 90.1\% of cases, while pairwise agreement among the LLM variants ranges from 93.6\% to 96.0\%. Agreement is typically very high for narrower and more explicitly signalled domains such as Patent, NHS, Drug Trial, Museum, Manufacturing, and Startup. Disagreement is concentrated in broader domains such as School, Software, Charity, Legislation, and Heritage, where case studies often describe indirect routes to impact or use more heterogeneous language. We therefore treat the GPT-5.5 classifications as the primary domain allocation for the regression and descriptive analyses, and use the regular-expression and alternative LLM classifications as cross-checks. Overall, the validation indicates that the allocation of case studies to domains is stable across reasonable classification strategies, with residual differences concentrated in conceptually broad categories rather than reflecting general instability in the domain-assignment procedure.

\subsection{Regression Models and Robustness}\label{sec:si-domain-analyses}

\subsubsection{Regression models}\label{sec:si-regression-modelling}

The regression analysis models variation in women's representation across impact case studies while incrementally accounting for disciplinary location, institutional characteristics, and impact domains. The unit of analysis is an impact case study \(i\) with at least one gender-identifiable contributor; all primary regressions are estimated on the 6,173 case studies meeting this criterion. Let \(W_i\) and \(M_i\) denote the recovered counts of women and men contributors, \(N_i=W_i+M_i\), and \(Y_i=W_i/N_i\), so that \(Y_i\) is the case-study share of gender-identifiable contributors who are women. The primary estimator is a weighted OLS regression of \(Y_i\), with analytic weight \(\omega_i=N_i\), so that case studies with more gender-identifiable contributors contribute proportionally more information. In all OLS equations, \(\alpha\) is the intercept and \(\varepsilon_i\) is the residual error term. Let \(P^B_i\), \(P^C_i\), and \(P^D_i\) be indicator variables equal to one when case study \(i\) is submitted to REF Panel B, Panel C, or Panel D, respectively; Panel~A is the omitted reference panel.\\

\noindent Model One estimates:
\[
Y_i=\alpha+\beta_B P^B_i+\beta_C P^C_i+\beta_D P^D_i+\varepsilon_i.
\]
Model Two adds institution-type controls. Let \(O_i\) indicate Oxbridge institutions, \(RG_i\) Russell Group institutions, \(RB_i\) Red Brick universities, and \(AN_i\) Ancient universities. Model Two estimates:
\[
Y_i=\alpha+\beta_B P^B_i+\beta_C P^C_i+\beta_D P^D_i+\gamma_O O_i+\gamma_{RG} RG_i+\gamma_{RB} RB_i+\gamma_{AN} AN_i+\varepsilon_i.
\]
The \(\gamma\) coefficients therefore capture the marginal association between each institutional category and \(Y_i\), conditional on panel and the other institutional indicators. These categories are not mutually exclusive: for example, an institution can be both Oxbridge and Russell Group.\\

Model Three adds impact-domain indicators. Let \(T_{ik}\) equal one when case study \(i\) is assigned to impact domain \(k\), where \(k=1,\ldots,K\) and \(K=11\) for Charity, Startup, Patent, Museum, NHS, Drug Trial, School, Legislation, Heritage, Manufacturing, and Software. Model Three estimates:
\[
Y_i=\alpha+\beta_B P^B_i+\beta_C P^C_i+\beta_D P^D_i+\gamma_O O_i+\gamma_{RG} RG_i+\gamma_{RB} RB_i+\gamma_{AN} AN_i+\sum_{k=1}^{K}\delta_k T_{ik}+\varepsilon_i.
\]
Domain indicators are also not mutually exclusive, so a case study can have multiple \(T_{ik}=1\). Each \(\delta_k\) compares case studies assigned to domain \(k\) with otherwise similar case studies not assigned to that domain, holding panel, institutional characteristics, and the other domain indicators fixed. These coefficients describe conditional associations and are not interpreted as causal effects. Weighted OLS coefficients are reported on the proportion scale and are interpreted as percentage-point differences after multiplication by 100. As a second estimator, we fit the same three specifications as a binomial generalised linear model with a logit link. In this specification,
\[
W_i\mid N_i\sim\mathrm{Binomial}(N_i,\pi_i),
\]
where \(\pi_i\) is the expected probability that a gender-identifiable contributor to case study \(i\) is a woman, and the log odds \(\log\{\pi_i/(1-\pi_i)\}\) are modelled using the same right-hand side as Model One, Model Two, or Model Three above. This estimator treats the numerator \(W_i\) and denominator \(N_i\) explicitly and constrains fitted values to the unit interval. The OLS models are used for the main coefficient plot because their coefficients are directly interpretable as changes in women's share; the GLM models provide a robustness check using the binomial likelihood and log-odds scale (Section~\ref{sec:si-glm}). Two further specifications build on Model Three. The supplementary UoA-control specification (Section~\ref{sec:si-uoa-controls}) keeps the same institutional and impact-domain structure but replaces the REF panel indicators with Unit of Assessment indicators, using Clinical Medicine as the reference category. For the Physics--Chemistry comparison (Section~\ref{sec:si-physics-chemistry}), the domain-predicted component for UoA \(u\) is computed from Model Three as \(\sum_k s_{uk}\hat{\delta}_k\), where \(s_{uk}\) is the share of case studies in UoA \(u\) assigned to domain \(k\).

\subsubsection{Domain composition and gender representation: physics vs chemistry}\label{sec:si-physics-chemistry}

Physics (UoA~9) and Chemistry (UoA~8) provide a focused comparison of how domain composition is associated with gender representation within two STEM fields strongly dominated by men. As shown in Table~\ref{tab:supp-uoa-summary}, women account for 14.05\% of gender-identifiable impact authorships and 6.94\% of output authorships in Physics, whereas in Chemistry they account for 19.11\% of impact authorships and 24.97\% of output authorships. Physics therefore shows a positive impact--output gap, while Chemistry shows a negative one. Table~\ref{tab:physics_chemistry_domains} contextualises this divergence by comparing the impact-domain mix of the two fields and attaching each domain to its estimated association with women's representation, using the Model Three coefficients reported in Table~\ref{tab:regressions}. Because the two fields differ in size (169 case studies in Physics, 113 in Chemistry), we compare domain shares rather than counts. Physics has a larger share of its case studies in domains associated with higher women's representation, including School (24.9\%, or 42 of 169 case studies, against 8.0\% in Chemistry), Museum (12.4\% against 4.4\%), and Charity (14.2\% against 12.4\%). Chemistry is instead more concentrated in domains associated with lower women's representation, especially Patent (61.9\%, or 70 of 113 case studies, against 33.7\% in Physics), Manufacturing (58.4\% against 42.6\%), and Drug Trial (30.1\% against
5.9\%). The corresponding Model Three OLS coefficients are \(-6.7\) percentage points for Patent (95\% CI, \(-9.8\) to \(-3.6\)), \(-4.7\) for Manufacturing (\(-7.4\) to \(-2.1\)), and \(-7.9\) for Drug Trial (\(-10.7\) to \(-5.1\)). The final rows of Table~\ref{tab:physics_chemistry_domains} summarise this domain mix by weighting the Model Three domain coefficients by the share of case studies in each UoA assigned to each domain, as specified in Section~\ref{sec:si-regression-modelling}. The resulting descriptive domain-predicted component is \(-2.9\) percentage points for Physics and \(-8.5\) percentage points for Chemistry under the OLS specification, with the parallel GLM summary also more negative for Chemistry (\(-0.51\) log-odds) than for Physics (\(-0.22\)). These are domain-share-weighted summaries of the point estimates, rather than additional fitted coefficients; 95\% CIs for the contributing OLS and GLM coefficients are shown in \figpanelref{fig:fig2}{b} and \figref{fig:supp_glm}, with standard errors in Table~\ref{tab:regressions}. The comparison illustrates that gender differences in impact vary with both the gender composition of fields and the types of impact activities those fields tend to produce.

\subsubsection{Alternative model specification}\label{sec:si-glm}
\figref{fig:supp_glm} reproduces the regression design from \figpanelref{fig:fig2}{b} using a binomial generalised linear model with a logit link, rather than the weighted OLS estimator used for the main coefficient plot. The OLS specification models the case-study share of gender-identifiable contributors who are women and has the advantage that coefficients can be read directly as percentage-point changes. The GLM instead models the count of women contributors out of the total number of gender-identifiable contributors in each case study, constraining fitted probabilities to the unit interval and expressing associations on the log-odds scale. Because the two estimators use different scales, the coefficients should not be compared numerically one-for-one; the relevant robustness question is whether the signs, ordering, and substantive interpretation of the associations are preserved. They are. Panel~B remains the strongest negative association with women's representation after institutional and domain controls are added: the Model Three GLM coefficient is \(-1.004\) (95\% CI, \(-1.122\) to \(-0.886\)) on the log-odds scale, matching in sign and relative magnitude the OLS coefficient of \(-20.0\) percentage points (95\% CI, \(-22.6\) to \(-17.4\)). Impact domains retain the same broad structure: Charity, NHS, School, and Museum are positively associated with women's representation in both estimators, while Patent, Drug Trial, Manufacturing, Startup, and Software are negatively associated. Institutional indicators remain weaker and less central than disciplinary and domain variables. The GLM results therefore show that the findings are not an artefact of using a linear probability-style model for a fractional outcome. The main OLS results are retained because they are easier to interpret in percentage-point terms, but the binomial logit specification supports the same substantive conclusion: gender composition varies systematically by disciplinary context and by the type of impact being produced.

\subsubsection{Robustness to finer discipline controls}\label{sec:si-uoa-controls}

\figref{fig:supp_uoa_models} examines whether the regression results are robust to a more detailed treatment of disciplinary composition. The main models use REF main panels because they provide a compact and interpretable control for broad disciplinary location, separating medicine and life sciences, physical sciences and engineering, social sciences, and arts and humanities. Panels are nevertheless coarse groupings, and each contains fields with distinct gender compositions, publication cultures, and routes to impact. This supplementary specification therefore replaces the panel indicators with Unit of Assessment indicators, using Clinical Medicine as the reference category, while retaining the same institutional controls and impact-domain indicators; the figure presents it using both the OLS and binomial GLM estimators. The UoA coefficients confirm substantial disciplinary heterogeneity beneath the panel averages: fields such as Physics, Computer Science and Informatics, Engineering, Economics and Econometrics, and Mathematical Sciences are among those most negatively associated with women's representation relative to Clinical Medicine, whereas several health, social science, education, and humanities UoAs are more positively positioned. The central question, however, is whether the domain coefficients disappear once this finer disciplinary structure is included. They do not. Domains such as Charity, Museum, NHS, and School remain positively associated with women's representation, while Patent, Drug Trial, Manufacturing, Software, and Startup remain negatively associated. Domain associations are therefore not simply proxies for broad panel composition or for a few highly gendered disciplines: the type of impact pathway still carries information about gender composition after controlling for much finer disciplinary location. The UoA-control models thus support the main interpretation; gender differences among named underpinning researchers are associated both with where researchers are located in the disciplinary structure and with the kinds of impact activity those fields produce.

\subsubsection{Robustness to impact-domain classification}\label{sec:si-domain-classification-robustness}
\figref{fig:supp_domain_models} assesses whether the domain-level gender patterns depend on the particular classification strategy used to assign impact case studies to domains. This is an important robustness check because the domain variables are inferred from unstructured case-study text rather than observed directly in REF metadata. The figure therefore reproduces the domain-by-panel descriptive pattern shown in \figpanelref{fig:fig2}{a} using alternative domain assignments: the deterministic regular-expression classifier and three alternative LLM-based classifications (Section~\ref{sec:si-impact-domains}). The purpose is not to treat any single alternative as a new preferred classification, but to test whether the substantive ordering of domains is sensitive to reasonable changes in how the same conceptual labels are operationalised. Across these specifications, the broad pattern is stable. Domains linked to patenting, manufacturing, startups, software, and drug trials remain among those with lower shares of women contributors, while domains linked to schools, museums, charities, the NHS, and heritage remain comparatively more gender-balanced. The panel structure is also preserved: Panel~B remains consistently lower than the other panels within many domains, indicating that the domain classification does not absorb all disciplinary differences in gender composition. Some variation across classifiers is expected, especially for broader domains such as Charity, School, Software, and Legislation, where case studies can describe indirect or diffuse pathways to impact --- consistent with the agreement patterns reported in Section~\ref{sec:si-impact-domain-validation} --- but this variation does not overturn the central descriptive result. The gendered structure of impact domains is therefore not an artefact of one prompt, one model version, or one keyword rule set: it appears in both rule-based and model-based classifications, supporting the interpretation that different forms of impact activity are associated with systematically different gender compositions. Finally, because some Panel~D assignments to \textit{Patent} and \textit{Drug Trial} may look surprising at first glance, Table~\ref{tab:panelD_unusual_domains} tabulates these low-frequency cases by Unit of Assessment, showing that they are rare and concentrated in a small number of UoAs.

\subsection{Qualitative Interviews}\label{sec:si-qualitative-interviews}

To contextualise the quantitative analyses presented in this study, we draw on qualitative interviews carried out between March and December 2023 for prior research on impact within the United Kingdom's Research Excellence Framework, as part of the work underpinning the British Academy report \textit{The SHAPE of Research Impact} \citesup{Wagner2024_sup}. The interviews covered two groups directly involved in the REF2021 impact assessment process: members of REF assessment panels responsible for evaluating impact case studies (\(N=36\)), sampled across the four main REF panels, and authors of impact case studies (\(N=12\)), interviewed to document how individual impact projects were developed and how researchers experienced the production and submission of case studies. All interviews followed a semi-structured format. Panel member interviews examined how impact was understood and evaluated within the REF, including assessment criteria and disciplinary differences; interviews with case study authors focused on the development of impact projects and the relationship between impact activities and broader research careers. The interviews were conducted by the wide research team at the University of Oxford's Leverhulme Centre for Demographic Science, lasted approximately 45 minutes, and were recorded and transcribed. The study received ethical approval from the University of Oxford (study number: SOC\_R2\_001\_C1A\_23\_03), and all participants provided informed consent. Panel member interviews were fully anonymised, while case study authors could choose whether to be quoted anonymously or by name. One case study author, Jo Appleby, is quoted by name in the main text; the quotations are verbatim from the interview recording, lightly edited for readability, and were reviewed and approved by Dr~Appleby in context, with written consent for named attribution in this article. The interviews were not originally designed to investigate gender differences in research impact. Nevertheless, gender-related themes --- including the compatibility of impact work with career interruptions, alternative routes to recognition, and differences in engagement activities across fields --- emerged repeatedly and unprompted, and motivated the quantitative analysis of gender and impact presented here. The interviews are particularly useful as contextual evidence because the quantitative analysis reveals not only an aggregate difference between impact and outputs but also substantial heterogeneity across the REF: for example, 1,301 case studies (20.45\% of all cases) have only women among gender-identifiable contributors, a share that ranges from 5.60\% in Panel~B to 33.18\% in Panel~D. We therefore use the interviews as supporting qualitative evidence to inform the interpretation of the patterns observed in the REF impact case study data, rather than as an independent analysis.

\subsection{Supplementary Descriptive Results}\label{sec:si-descriptive-results}

\subsubsection{Descriptive statistics by REF main panel}\label{sec:si-panel-descriptives}

Table~\ref{tab:supp-panel-summary} reports the panel-level descriptive statistics underlying \figref{fig:gender_ratios}. It summarises the four REF main panels, Panel~A (medicine and life sciences), Panel~B (physical sciences and engineering), Panel~C (social sciences), and Panel~D (arts and humanities) by reporting measures of submitted staff full-time equivalent (FTE), doctoral degrees awarded, total research income, the number of impact case studies, the number of submitted research outputs, and the corresponding shares of women among gender-identifiable contributors. The panels differ substantially in scale, which matters for interpretation because aggregate figures weight the panels unevenly. Panel~C is the largest by submitted staff, with 23,451 FTE, while Panel~A is largest by research income and submitted outputs, with \pounds 24.25 billion in total research income and 48,555 submitted outputs. Panel~C also contributes the largest number of impact case studies (\(N=2,146\)). Panel~D is smaller on these measures, with 14,305 FTE and 16,315 submitted outputs, but contains 1,528 impact case studies. Women's representation varies sharply across panels: women account for 45.51\% of gender-identifiable impact authorships in Panel~A, 18.27\% in Panel~B, 41.57\% in Panel~C, and 46.39\% in Panel~D. The same broad ordering is visible in research outputs, where women account for 40.02\% of authorships in Panel~A, 15.92\% in Panel~B, 36.17\% in Panel~C, and 44.62\% in Panel~D. The impact share is higher than the output share in every panel: by 5.49 percentage points in Panel~A, 5.40 in Panel~C, 2.35 in Panel~B, and 1.77 in Panel~D. The panel-level descriptives therefore support the central finding that women are more represented in impact than in submitted outputs, while also showing that this pattern is layered on top of substantial disciplinary stratification.

\subsubsection{Descriptive statistics and representation of women by unit of assessment}\label{sec:si-uoa-descriptives}
Table~\ref{tab:supp-uoa-summary} reports the same descriptive measures as Table~\ref{tab:supp-panel-summary}, but disaggregated across the 34 REF Units of Assessment (UoAs): submitted staff FTE, doctoral degrees awarded, research income, numbers of impact case studies and research outputs, and the corresponding shares of women among gender-identifiable contributors. These values provide the UoA-level basis for the patterns shown in \figref{fig:gender_ratios} and show that panel-level averages conceal substantial disciplinary heterogeneity. Women's representation in impact authorships is lowest in Physics (14.05\%), Engineering (16.06\%), and Computer Science and Informatics (16.22\%), and highest in Social Work and Social Policy (57.30\%), English Language and Literature (57.17\%), and Allied Health Professions, Dentistry, Nursing and Pharmacy (57.15\%). The impact--output gap also varies considerably across fields. The largest positive percentage-point gaps are in Classics (+14.14), Allied Health Professions, Dentistry, Nursing and Pharmacy (+13.63), and Psychology, Psychiatry and Neuroscience (+12.55). By contrast, the largest negative gaps are in Clinical Medicine (-8.59), Biological Sciences (-7.31), and Chemistry (-5.86). These UoA-level contrasts make clear that the overall impact advantage for women is not uniform across disciplines, but is concentrated in particular areas of the REF. Note that these percentage-point gaps order fields differently from the ratios shown in \figpanelref{fig:gender_ratios}{d}, which express women's and men's representation in impact relative to their representation in outputs: because a given absolute gap is
proportionally larger where baseline representation is low, fields such
as Physics show the largest relative gains.

\subsubsection{Descriptive statistics and gender representation across impact domains}\label{sec:si-domain-descriptives}

Tables~\ref{tab:supp-llm-summary} and \ref{tab:supp-llm-panel-summary} report the descriptive statistics underlying the impact-domain patterns shown in \figpanelref{fig:fig2}{a}. Table~\ref{tab:supp-llm-summary} gives the overall number of impact case studies assigned to each domain and the share of gender-identifiable contributors in those case studies who are women; Table~\ref{tab:supp-llm-panel-summary} provides the same quantities separately by REF main panel. Because domain assignment is multi-label, a case study can contribute to more than one domain: the domain counts are not mutually exclusive, and the column reporting the percentage of all women authors gives the share of gender-identifiable women impact authorships appearing in case studies assigned to that domain, not a partition of women authorships across domains. The most prevalent domains are Charity (\(N=2,595\)), Software (\(N=1,848\)), Legislation (\(N=1,448\)), School (\(N=1,263\)), and NHS (\(N=1,257\)). Women's representation is highest in Charity (47.54\%), School (47.04\%), Museum (46.49\%), and NHS (45.45\%), and lowest in Patent (17.89\%), Manufacturing (20.02\%), Startup (22.83\%), Software (32.64\%), and Drug Trial (33.26\%). The panel-specific breakdown shows that these domain differences are layered on top of disciplinary stratification. Panel~B is heavily represented in Software (\(N=701\)), Manufacturing (\(N=439\)), and Startup (\(N=335\)), all of which have low women's representation within that panel; Panel~D, by contrast, is concentrated in Museum (\(N=687\)), Charity (\(N=685\)), Heritage (\(N=618\)), and School (\(N=546\)), where women's representation is close to parity. These tables therefore show both the prevalence of different impact pathways and the gender composition of the contributors associated with each pathway.

\subsubsection{Text-based analysis of gender differences in impact case studies}\label{sec:si-text-analysis}

\begin{figure}[!ht]\centering
\includegraphics[width=0.75\textwidth]{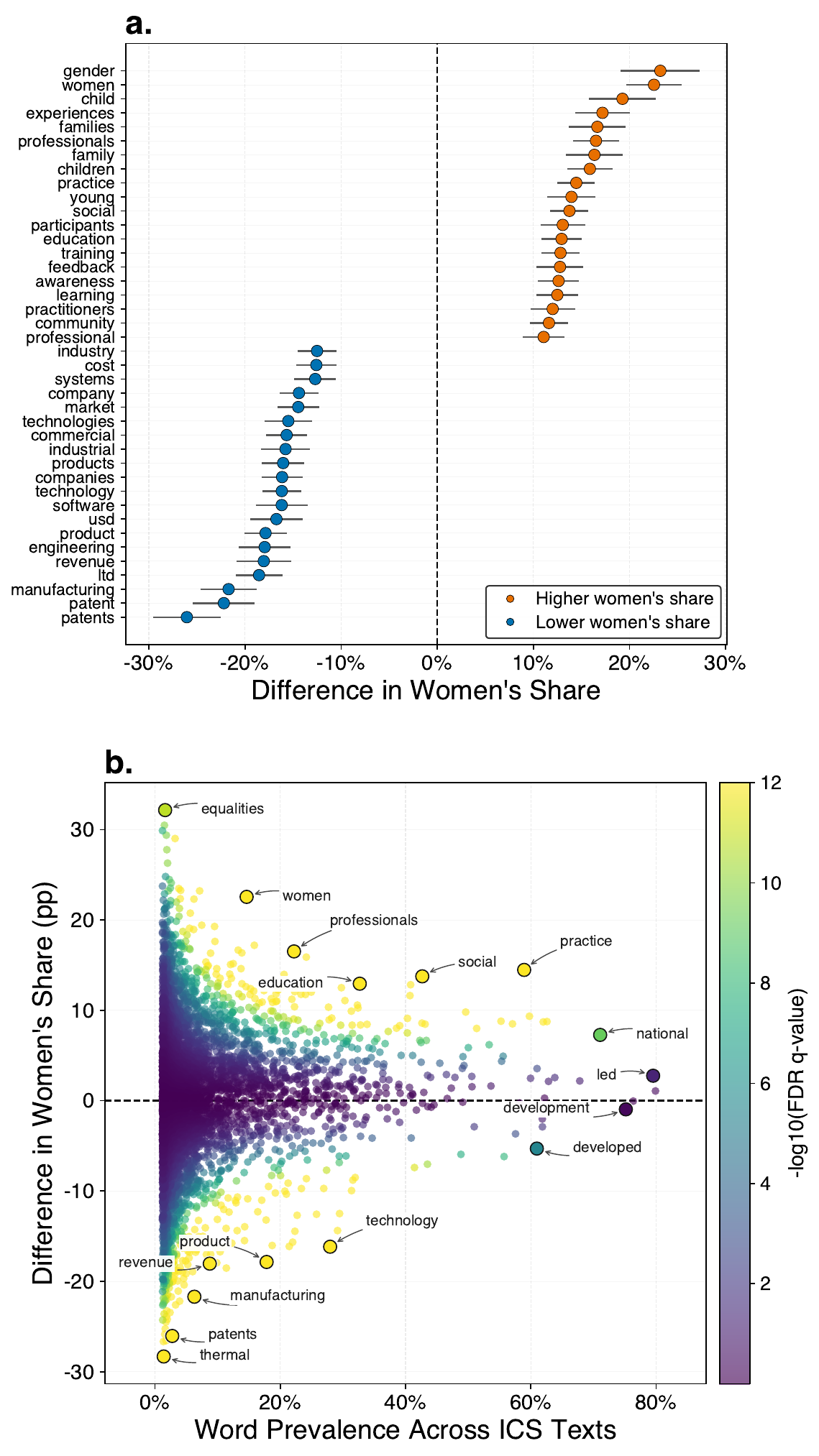}
\caption{
\textbf{Associations between word usage in impact case studies and women’s representation.}
Subfigure `\textbf{a.}' shows words associated with higher and lower shares of women authors. Subfigure `\textbf{b.}' relates word prevalence to differences in women’s representation, with statistical significance indicated by false discovery rate.
}
\label{fig:supp_text}
\end{figure}

\begin{figure}[!ht]\centering
\includegraphics[width=\textwidth]{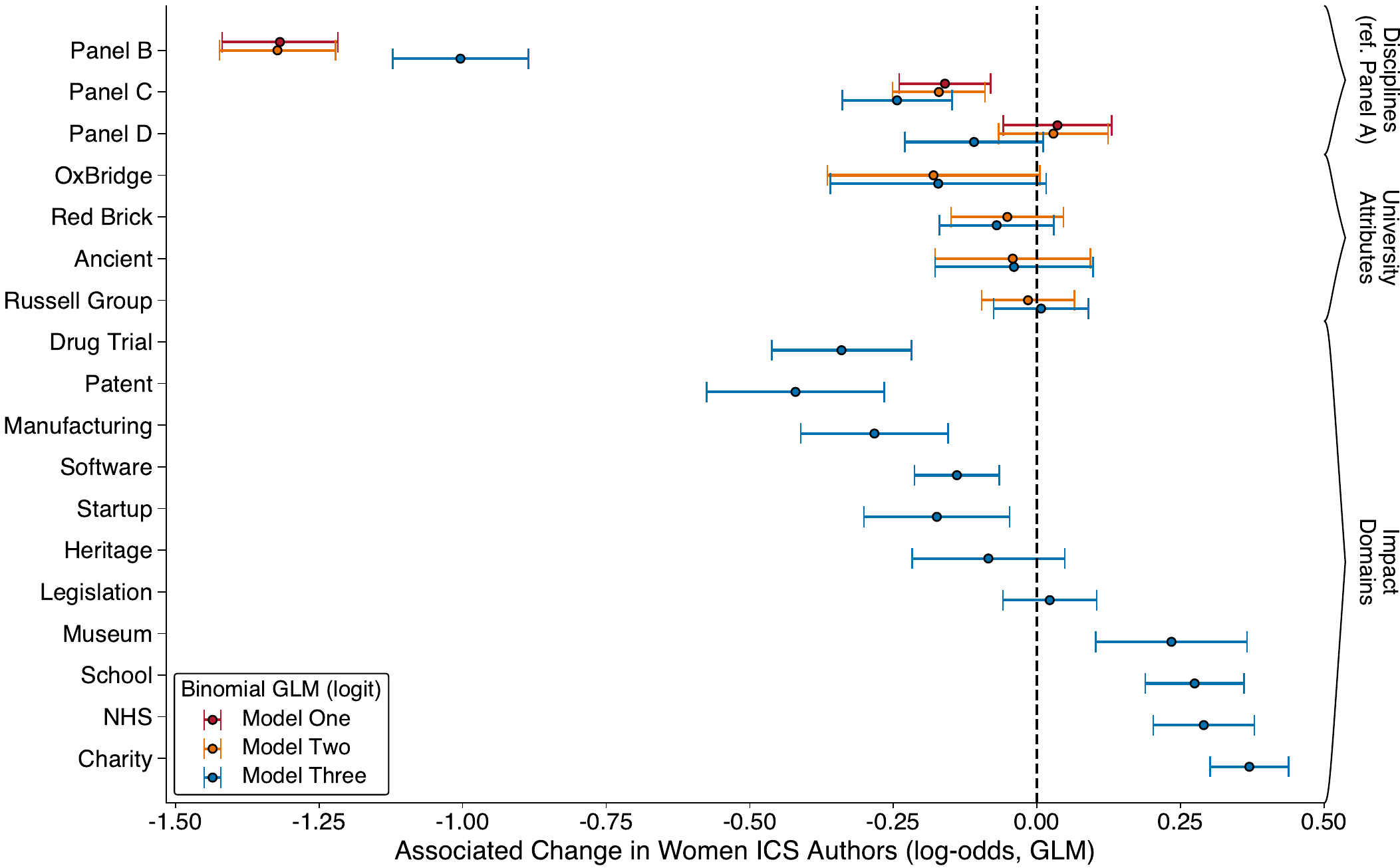}
\caption{
\textbf{Robustness of model estimates to alternative estimation strategy.}
This figure reproduces the model shown in \figpanelref{fig:fig2}{b}, but instead here using a binomial GLM (logit) estimator.
}
\label{fig:supp_glm}
\end{figure}

\begin{figure}[!ht]\centering
\includegraphics[width=\textwidth]{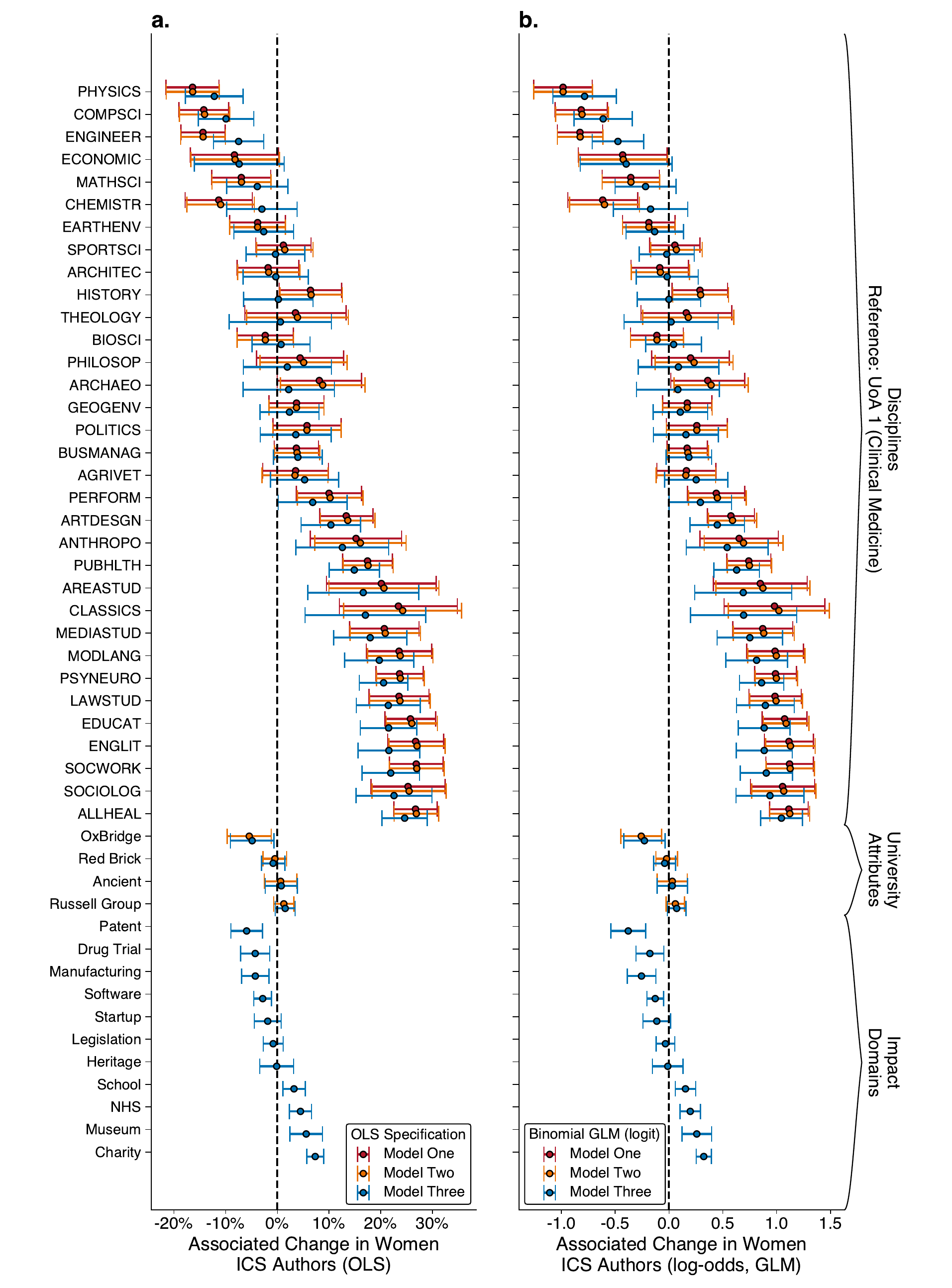}
\caption{
\textbf{Robustness of model estimates to finer disciplinary controls.} Subfigure `\textbf{a.}' shows a finer, more granular estimation of `Model Three' from \figpanelref{fig:fig2}{b} which replaces panels with Units of Assessment, again estimated with OLS. Subfigure `\textbf{b.}' does the same thing, but estimates the model with a generalised linear model akin to Figure \ref{fig:supp_glm}.}
\label{fig:supp_uoa_models}
\end{figure}

\begin{figure}[!ht]\centering
\includegraphics[width=\textwidth]{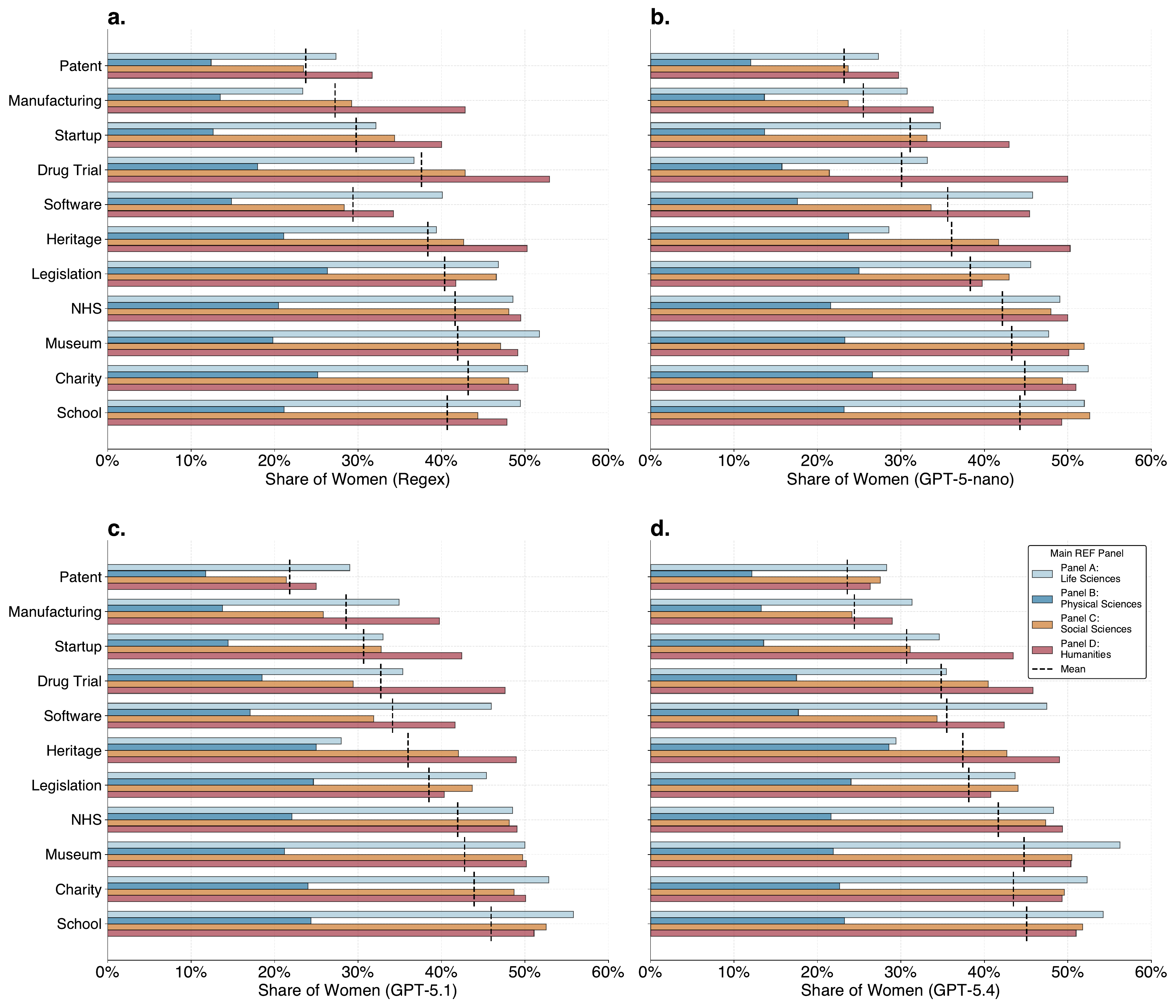}
\caption{
\textbf{Robustness of gender patterns across impact domains to alternative classification approaches.}
Subfigure `\textbf{a.}' shows domain-level patterns with domains covered by regular expressions. Subfigures `\textbf{b.}'-`\textbf{d.}' show alternate large language modelling based approaches, equivalent to \figpanelref{fig:fig2}{a}.
}
\label{fig:supp_domain_models}
\end{figure}

\figref{fig:supp_text} provides a data-driven, text-based check on the domain-classification results. Instead of assigning each case study to a predefined set of impact domains, this analysis asks whether individual terms in the impact narratives are associated with higher or lower shares of women contributors. The analysis uses the substantive narrative sections of each case study, constructs a vocabulary of terms that appear often enough to support comparison, and estimates the difference in mean women's representation between case studies where each term is present and case studies where it is absent, applying a false-discovery-rate correction to identify statistically robust associations. This approach is deliberately less structured than the LLM domain classification: it does not know in advance which words should correspond to Charity, Patent, School, Manufacturing, or any other domain. Its value is therefore diagnostic. If the domain findings were mainly a product of the classification scheme, the free-text analysis would not necessarily recover the same pattern. In practice, it does. Terms associated with higher shares of women contributors include words linked to social, educational, professional, and community-oriented impact, such as ``practice'', ``social'', ``professionals'', ``children'', ``education'', ``community'', and ``learning''. Terms associated with lower shares of women contributors are concentrated around commercial, industrial, technological, and market-oriented pathways, including ``technology'', ``product'', ``companies'', ``commercial'', ``manufacturing'', ``patent'', ``industry'', and ``software''. These word-level associations closely mirror the domain-level results, showing that the same gendered structure emerges directly from the language of the case studies, without requiring the predefined domain labels; the text analysis therefore provides an independent validation of the domain-based interpretation.

\subsection{Abbreviations and Figure Labels}\label{sec:si-abbreviations}

Table~\ref{tab:supp-abbreviations} lists the abbreviations used in the main text and supplementary materials. Throughout, Panels A--D denote the four REF main panels: Panel~A (medicine, health and life sciences), Panel~B (physical sciences, engineering and mathematics), Panel~C (social sciences), and Panel~D (arts and humanities). Because the full REF2021 Unit of Assessment names are too long for figure axes, Table~\ref{tab:supp-uoa-labels} maps the shortened labels used in the figures to the official UoA names; institution-type and impact-domain labels appear in full in all figures.

\clearpage
\subsection*{Supplementary Figures}

\newpage

\newpage

\newpage

\newpage

\newpage

\subsection*{Supplementary Tables}

\begin{table}[htbp]
\centering
\caption{Abbreviations used in the main text and supplementary materials.}
\label{tab:supp-abbreviations}
\small
\begin{tabular}{ll}
\toprule
Abbreviation & Full term \\
\midrule
API     & Application Programming Interface \\
DOI     & Digital Object Identifier \\
FDR     & False Discovery Rate \\
FTE     & Full-Time Equivalent (submitted staff) \\
GLM     & Generalised Linear Model \\
ICS     & Impact Case Study \\
ISBN    & International Standard Book Number \\
JSON    & JavaScript Object Notation \\
LLM     & Large Language Model \\
NGO     & Non-Governmental Organisation \\
NHS     & National Health Service \\
OLS     & Ordinary Least Squares \\
REF     & Research Excellence Framework \\
REF2021 & The 2021 exercise of the Research Excellence Framework \\
STEM    & Science, Technology, Engineering and Medicine \\
UK      & United Kingdom \\
UKPRN   & UK Provider Reference Number \\
UoA     & Unit of Assessment \\
\bottomrule
\end{tabular}
\end{table}

\newpage

\begin{table}[htbp]
\centering
\caption{Shortened Unit of Assessment labels used in the figures and the
corresponding official REF2021 Unit of Assessment names.}
\label{tab:supp-uoa-labels}
\scriptsize
\begin{tabular}{rllc}
\toprule
UoA & Official REF2021 name & Figure label & Panel \\
\midrule
1  & Clinical Medicine & Clinical Medicine & A \\
2  & Public Health, Health Services and Primary Care & Public Health & A \\
3  & Allied Health Professions, Dentistry, Nursing and Pharmacy & Allied Health Prof. & A \\
4  & Psychology, Psychiatry and Neuroscience & Psychology \& Neuro. & A \\
5  & Biological Sciences & Biological Sciences & A \\
6  & Agriculture, Food and Veterinary Sciences & Agri. \& Vet. Sciences & A \\
7  & Earth Systems and Environmental Sciences & Earth \& Env. Sciences & B \\
8  & Chemistry & Chemistry & B \\
9  & Physics & Physics & B \\
10 & Mathematical Sciences & Mathematical Sciences & B \\
11 & Computer Science and Informatics & Comp. Sci. \& Informatics & B \\
12 & Engineering & Engineering & B \\
13 & Architecture, Built Environment and Planning & Arch. \& Built Env. & C \\
14 & Geography and Environmental Studies & Geog. \& Env. Studies & C \\
15 & Archaeology & Archaeology & C \\
16 & Economics and Econometrics & Economics & C \\
17 & Business and Management Studies & Business \& Management & C \\
18 & Law & Law & C \\
19 & Politics and International Studies & Politics \& Int. Studies & C \\
20 & Social Work and Social Policy & Social Work \& Policy & C \\
21 & Sociology & Sociology & C \\
22 & Anthropology and Development Studies & Anthro. \& Dev. Studies & C \\
23 & Education & Education & C \\
24 & Sport and Exercise Sciences, Leisure and Tourism & Sport \& Exercise Sci. & C \\
25 & Area Studies & Area Studies & D \\
26 & Modern Languages and Linguistics & Modern Lang. \& Ling. & D \\
27 & English Language and Literature & English Lang. \& Lit. & D \\
28 & History & History & D \\
29 & Classics & Classics & D \\
30 & Philosophy & Philosophy & D \\
31 & Theology and Religious Studies & Theology \& Religion & D \\
32 & Art and Design: History, Practice and Theory & Art \& Design & D \\
33 & Music, Drama, Dance, Performing Arts, Film and Screen Studies & Music, Drama \& Perf. Arts & D \\
34 & Communication, Cultural and Media Studies, Library and Information Management & Communication \& Media & D \\
\bottomrule
\end{tabular}
\end{table}

\newpage

\begin{table}[!ht]\centering
\caption{Distribution of impact case studies and representation of women across impact domains.}
\label{tab:supp-llm-summary}
\footnotesize
\setlength{\tabcolsep}{3pt}
\renewcommand{\arraystretch}{0.95}
\begin{adjustbox}{max width=\textwidth}
\begin{tabular}{lccc}
\toprule
Impact domain & Number of ICS & \% Women Authors & \% of All Women Authors \\
\midrule
Charity & 2595 & 47.54 & 50.85 \\
Startup & 638 & 22.83 & 7.09 \\
Patent & 516 & 17.89 & 4.49 \\
Museum & 904 & 46.49 & 13.93 \\
NHS & 1257 & 45.45 & 29.96 \\
Drug Trial & 558 & 33.26 & 9.78 \\
School & 1263 & 47.04 & 23.94 \\
Legislation & 1448 & 39.17 & 22.71 \\
Heritage & 851 & 44.14 & 12.24 \\
Manufacturing & 695 & 20.02 & 6.58 \\
Software & 1848 & 32.64 & 28.80 \\
\bottomrule
\end{tabular}

\end{adjustbox}
\end{table}

\newpage

\begin{table}[!ht]\centering
\caption{Distribution of impact domains and representation of women within domains by REF main panel.}
\label{tab:supp-llm-panel-summary}
\footnotesize
\setlength{\tabcolsep}{3pt}
\renewcommand{\arraystretch}{0.95}
\begin{adjustbox}{max width=\textwidth}
\begin{tabular}{llccc}
\toprule
Panel & Impact domain & Number of ICS & \% Women Authors & \% of All Women Authors (panel) \\
\midrule
A & Charity & 646 & 51.52 & 54.31 \\
A & Startup & 150 & 32.89 & 6.98 \\
A & Patent & 164 & 27.74 & 6.32 \\
A & Museum & 23 & 50.75 & 1.60 \\
A & NHS & 758 & 48.73 & 63.32 \\
A & Drug Trial & 409 & 35.71 & 24.05 \\
A & School & 159 & 53.29 & 11.46 \\
A & Legislation & 306 & 42.86 & 19.24 \\
A & Heritage & 26 & 33.82 & 1.08 \\
A & Manufacturing & 136 & 34.14 & 6.65 \\
A & Software & 381 & 46.82 & 30.17 \\
B & Charity & 210 & 25.30 & 24.67 \\
B & Startup & 335 & 13.89 & 21.04 \\
B & Patent & 308 & 12.01 & 16.84 \\
B & Museum & 66 & 23.41 & 8.56 \\
B & NHS & 140 & 21.75 & 14.08 \\
B & Drug Trial & 116 & 20.71 & 10.16 \\
B & School & 128 & 23.65 & 17.13 \\
B & Legislation & 226 & 22.09 & 21.48 \\
B & Heritage & 47 & 25.29 & 6.39 \\
B & Manufacturing & 439 & 13.56 & 25.83 \\
B & Software & 701 & 17.52 & 53.56 \\
C & Charity & 1054 & 49.30 & 58.10 \\
C & Startup & 98 & 29.64 & 4.19 \\
C & Patent & 29 & 22.47 & 0.92 \\
C & Museum & 128 & 50.77 & 7.59 \\
C & NHS & 240 & 46.02 & 14.35 \\
C & Drug Trial & 21 & 39.58 & 0.87 \\
C & School & 430 & 52.17 & 29.30 \\
C & Legislation & 761 & 43.07 & 34.04 \\
C & Heritage & 160 & 43.44 & 7.77 \\
C & Manufacturing & 87 & 25.39 & 2.99 \\
C & Software & 448 & 35.59 & 21.25 \\
D & Charity & 685 & 48.79 & 46.94 \\
D & Startup & 55 & 41.22 & 4.73 \\
D & Patent & 15 & 31.58 & 0.93 \\
D & Museum & 687 & 49.84 & 47.71 \\
D & NHS & 119 & 50.00 & 9.91 \\
D & Drug Trial & 12 & 46.88 & 1.16 \\
D & School & 546 & 49.56 & 39.04 \\
D & Legislation & 155 & 43.29 & 9.99 \\
D & Heritage & 618 & 47.97 & 41.21 \\
D & Manufacturing & 33 & 35.80 & 2.25 \\
D & Software & 318 & 43.92 & 26.03 \\
\bottomrule
\end{tabular}

\end{adjustbox}
\end{table}

\newpage

\begin{landscape}
\begin{table}[!ht]\centering
\caption{Descriptive statistics and representation of women in impact case studies and research outputs by Unit of Assessment.}
\label{tab:supp-uoa-summary}
\footnotesize
\setlength{\tabcolsep}{3pt}
\renewcommand{\arraystretch}{0.9}
\begin{adjustbox}{max width=\linewidth}
\begin{tabular}{rllrrrrrrrrr}
\toprule
UoA & Unit of Assessment & Panel & FTE & PhDs (000) & Total Income (£bn) & N (ICS) & \% Women (ICS) & \% All Women (ICS) & N (Papers) & \% Women (Papers) & \% All Women (Papers) \\
\midrule
1 & Clinical Medicine & A & 4878 & 12 & 11.01 & 254 & 30.44 & 4.25 & 11919 & 39.03 & 26.39 \\
2 & Public Health, Health Services and Primary Care & A & 2032 & 3 & 3.39 & 151 & 47.88 & 5.58 & 4908 & 41.25 & 11.51 \\
3 & Allied Health Professions, Dentistry, Nursing and Pharmacy & A & 4769 & 9 & 1.76 & 393 & 57.15 & 12.29 & 11527 & 43.52 & 12.82 \\
4 & Psychology, Psychiatry and Neuroscience & A & 4040 & 11 & 2.94 & 326 & 54.07 & 7.83 & 9697 & 41.53 & 10.83 \\
5 & Biological Sciences & A & 2867 & 9 & 4.00 & 192 & 28.10 & 2.17 & 7085 & 35.41 & 6.54 \\
6 & Agriculture, Food and Veterinary Sciences & A & 1398 & 3 & 1.14 & 103 & 33.97 & 1.69 & 3419 & 36.67 & 2.73 \\
7 & Earth Systems and Environmental Sciences & B & 1782 & 4 & 1.33 & 148 & 26.63 & 2.15 & 4364 & 27.53 & 2.63 \\
8 & Chemistry & B & 1502 & 7 & 1.82 & 113 & 19.11 & 0.89 & 3685 & 24.97 & 1.70 \\
9 & Physics & B & 2214 & 6 & 3.78 & 169 & 14.05 & 1.34 & 5191 & 6.94 & 1.93 \\
10 & Mathematical Sciences & B & 2461 & 5 & 0.70 & 176 & 23.47 & 1.59 & 5782 & 16.18 & 0.79 \\
11 & Computer Science and Informatics & B & 3002 & 7 & 1.37 & 271 & 16.22 & 1.91 & 6895 & 18.62 & 1.20 \\
12 & Engineering & B & 7432 & 24 & 7.43 & 391 & 16.06 & 3.09 & 18068 & 19.17 & 3.46 \\
13 & Architecture, Built Environment and Planning & C & 1527 & 3 & 0.34 & 127 & 28.65 & 1.69 & 3136 & 29.10 & 0.67 \\
14 & Geography and Environmental Studies & C & 1855 & 3 & 0.59 & 180 & 34.15 & 2.88 & 4243 & 29.41 & 2.11 \\
15 & Archaeology & C & 497 & 1 & 0.21 & 59 & 38.55 & 1.02 & 854 & 37.21 & 0.59 \\
16 & Economics and Econometrics & C & 920 & 1 & 0.19 & 88 & 22.15 & 0.53 & 2128 & 20.38 & 0.26 \\
17 & Business and Management Studies & C & 6634 & 9 & 0.53 & 504 & 34.12 & 6.47 & 15596 & 31.12 & 3.33 \\
18 & Law & C & 2494 & 3 & 0.17 & 226 & 53.96 & 3.47 & 3380 & 45.88 & 0.58 \\
19 & Politics and International Studies & C & 1962 & 3 & 0.26 & 166 & 36.17 & 1.62 & 3867 & 33.56 & 0.60 \\
20 & Social Work and Social Policy & C & 2105 & 3 & 0.37 & 222 & 57.30 & 5.13 & 4307 & 54.84 & 1.79 \\
21 & Sociology & C & 1104 & 2 & 0.27 & 107 & 55.70 & 2.02 & 2153 & 50.75 & 0.61 \\
22 & Anthropology and Development Studies & C & 733 & 2 & 0.22 & 77 & 45.65 & 1.00 & 1364 & 40.99 & 0.38 \\
23 & Education & C & 2168 & 6 & 0.39 & 230 & 56.17 & 6.17 & 4379 & 55.37 & 1.71 \\
24 & Sport and Exercise Sciences, Leisure and Tourism & C & 1453 & 2 & 0.15 & 160 & 31.61 & 2.63 & 3457 & 34.04 & 2.02 \\
25 & Area Studies & D & 580 & 1 & 0.08 & 57 & 50.55 & 0.73 & 918 & 45.74 & 0.17 \\
26 & Modern Languages and Linguistics & D & 1615 & 3 & 0.17 & 154 & 53.97 & 2.71 & 2031 & 55.53 & 0.49 \\
27 & English Language and Literature & D & 2671 & 5 & 0.14 & 273 & 57.17 & 4.57 & 2596 & 51.55 & 0.42 \\
28 & History & D & 2360 & 4 & 0.25 & 240 & 36.84 & 2.12 & 3202 & 38.15 & 0.37 \\
29 & Classics & D & 448 & 1 & 0.06 & 48 & 53.85 & 0.67 & 426 & 39.71 & 0.06 \\
30 & Philosophy & D & 692 & 1 & 0.10 & 85 & 34.84 & 0.86 & 1267 & 25.08 & 0.10 \\
31 & Theology and Religious Studies & D & 505 & 2 & 0.06 & 68 & 33.94 & 0.59 & 561 & 32.81 & 0.06 \\
32 & Art and Design: History, Practice and Theory & D & 2607 & 3 & 0.25 & 262 & 43.76 & 4.08 & 2104 & 44.97 & 0.53 \\
33 & Music, Drama, Dance, Performing Arts, Film and Screen Studies & D & 1523 & 3 & 0.10 & 196 & 40.44 & 2.06 & 1407 & 43.88 & 0.24 \\
34 & Communication, Cultural and Media Studies, Library and Information Management & D & 1303 & 2 & 0.10 & 145 & 51.12 & 2.18 & 1803 & 46.16 & 0.38 \\
\bottomrule
\end{tabular}

\end{adjustbox}
\end{table}
\end{landscape}

\newpage

\begin{table}[!ht]\centering
\caption{Distribution of impact domains in Physics (UoA 9) and Chemistry (UoA 8), with associated OLS and GLM estimates from Model Three.}
\label{tab:physics_chemistry_domains}
\footnotesize
\setlength{\tabcolsep}{3pt}
\renewcommand{\arraystretch}{0.95}
\begin{adjustbox}{max width=\textwidth}
\begin{tabular}{lcccc}
\toprule
Impact domain & Physics (UoA 9) & Chemistry (UoA 8) & Delta Women (pp, OLS) & Delta Women (log-odds, GLM) \\
\midrule
Charity & 24 & 14 & +8.7 & +0.37 \\
NHS & 15 & 12 & +6.8 & +0.29 \\
School & 42 & 9 & +6.3 & +0.27 \\
Museum & 21 & 5 & +5.1 & +0.23 \\
Legislation & 19 & 17 & +0.3 & +0.02 \\
Heritage & 6 & 3 & -1.8 & -0.08 \\
Startup & 56 & 46 & -3.0 & -0.17 \\
Software & 63 & 17 & -3.2 & -0.14 \\
Manufacturing & 72 & 66 & -4.7 & -0.28 \\
Patent & 57 & 70 & -6.7 & -0.42 \\
Drug Trial & 10 & 34 & -7.9 & -0.34 \\
\midrule
\textbf{Total ICS (UoA)} & \textbf{169} & \textbf{113} &  &  \\
\textbf{Domain predicted $\Delta$ Women (pp, OLS):} & \textbf{-2.9} & \textbf{-8.5} &  &  \\
\textbf{Domain predicted $\Delta$ log-odds (GLM):} & \textbf{-0.22} & \textbf{-0.51} &  &  \\
\bottomrule
\end{tabular}

\end{adjustbox}
\end{table}

\newpage

\begin{table}[!ht]\centering
\caption{Full regression results for models of women's representation in impact case studies.}
\label{tab:regressions}
\begin{center}
\begin{tabular}{lllllll}
\hline
                   & OLS (1)   & OLS (2)   & OLS (3)   & GLM (1)   & GLM (2)   & GLM (3)    \\
\hline
Panel B  & -0.272*** & -0.273*** & -0.200*** & -1.318*** & -1.322*** & -1.004***  \\
                   & (0.011)   & (0.011)   & (0.013)   & (0.051)   & (0.051)   & (0.060)    \\
Panel C  & -0.039*** & -0.042*** & -0.055*** & -0.160*** & -0.171*** & -0.244***  \\
                   & (0.010)   & (0.010)   & (0.012)   & (0.041)   & (0.041)   & (0.049)    \\
Panel D  & 0.009     & 0.007     & -0.021    & 0.036     & 0.029     & -0.110*    \\
                   & (0.012)   & (0.012)   & (0.015)   & (0.048)   & (0.049)   & (0.062)    \\
OxBridge           &           & -0.040*   & -0.038*   &           & -0.180*   & -0.172*    \\
                   &           & (0.023)   & (0.022)   &           & (0.094)   & (0.096)    \\
Russell Group       &           & -0.004    & 0.002     &           & -0.016    & 0.007      \\
                   &           & (0.010)   & (0.010)   &           & (0.041)   & (0.042)    \\
Red Brick           &           & -0.011    & -0.016    &           & -0.052    & -0.070     \\
                   &           & (0.012)   & (0.012)   &           & (0.050)   & (0.051)    \\
Ancient            &           & -0.009    & -0.008    &           & -0.042    & -0.040     \\
                   &           & (0.017)   & (0.016)   &           & (0.069)   & (0.070)    \\
llm\_museum        &           &           & 0.051***  &           &           & 0.234***   \\
                   &           &           & (0.016)   &           &           & (0.067)    \\
llm\_nhs           &           &           & 0.068***  &           &           & 0.290***   \\
                   &           &           & (0.011)   &           &           & (0.045)    \\
llm\_drug\_trial   &           &           & -0.079*** &           &           & -0.340***  \\
                   &           &           & (0.014)   &           &           & (0.062)    \\
llm\_school        &           &           & 0.063***  &           &           & 0.274***   \\
                   &           &           & (0.011)   &           &           & (0.044)    \\
llm\_legislation   &           &           & 0.003     &           &           & 0.022      \\
                   &           &           & (0.010)   &           &           & (0.042)    \\
llm\_heritage      &           &           & -0.018    &           &           & -0.085     \\
                   &           &           & (0.016)   &           &           & (0.068)    \\
llm\_manufacturing &           &           & -0.047*** &           &           & -0.283***  \\
                   &           &           & (0.013)   &           &           & (0.065)    \\
llm\_software      &           &           & -0.032*** &           &           & -0.139***  \\
                   &           &           & (0.009)   &           &           & (0.038)    \\
llm\_patent        &           &           & -0.067*** &           &           & -0.420***  \\
                   &           &           & (0.016)   &           &           & (0.079)    \\
llm\_startup       &           &           & -0.030**  &           &           & -0.174***  \\
                   &           &           & (0.014)   &           &           & (0.065)    \\
llm\_charity       &           &           & 0.087***  &           &           & 0.370***   \\
                   &           &           & (0.008)   &           &           & (0.035)    \\
Intercept          & 0.455***  & 0.464***  & 0.420***  & -0.180*** & -0.142*** & -0.313***  \\
                   & (0.008)   & (0.009)   & (0.012)   & (0.029)   & (0.035)   & (0.052)    \\
R-squared          & 0.108     & 0.110     & 0.164     &           &           &            \\
R-squared Adj.     & 0.108     & 0.109     & 0.162     &           &           &            \\
AIC                & 4857.4    & 4855.3    & 4488.4    & 17001.4   & 16997.4   & 16557.0    \\
Adj. R2            & 0.108     & 0.109     & 0.162     &           &           &            \\
BIC                & 4884.3    & 4909.1    & 4616.3    & -150465.3 & -150438.5 & -150794.1  \\
LogLik             & -2424.7   & -2419.6   & -2225.2   & -8496.7   & -8490.7   & -8259.5    \\
N                  & 6173      & 6173      & 6173      & 6173      & 6173      & 6173       \\
Pseudo R2          &           &           &           &           &           &            \\
R-squared          & 0.108     & 0.110     & 0.164     &           &           &            \\
\hline
\end{tabular}
\par\smallskip
\parbox{0.95\linewidth}{\footnotesize\raggedright\textit{Notes:} Standard errors in parentheses. $^{*}p<0.10$, $^{**}p<0.05$, $^{***}p<0.01$.}
\end{center}

\end{table}

\newpage

\begin{table}[!ht]\centering
\caption{Distribution of Panel~D case studies classified as \textit{Patent} and \textit{Drug Trial} by Unit of Assessment (LLM classification).}
\label{tab:panelD_unusual_domains}
\small
\begin{tabular}{lcc}
\toprule
UoA & Patent & Drug Trial \\
\midrule
32 Art and Design: History, Practice and Theory & 11 & 5 \\
30 Philosophy & 0 & 4 \\
34 Communication, Cultural and Media Studies & 2 & 1 \\
26 Modern Languages and Linguistics & 1 & 0 \\
27 English Language and Literature & 0 & 1 \\
31 Theology and Religious Studies & 0 & 1 \\
33 Performing Arts & 1 & 0 \\
\midrule
Total & 15 & 12 \\
\bottomrule
\end{tabular}

\end{table}

\newpage

\begin{table}[!ht]\centering
\caption{Descriptive statistics and representation of women in impact case studies and research outputs by REF main panel.}
\label{tab:supp-panel-summary}
\small
\setlength{\tabcolsep}{4pt}
\begin{adjustbox}{max width=\textwidth}
\begin{tabular}{lrrrrrrrrr}
\toprule
Panel & FTE & PhDs (000) & Total Income (£bn) & N (ICS) & \% Women (ICS) & \% All Women (ICS) & N (Papers) & \% Women (Papers) & \% All Women (Papers) \\
\midrule
A & 19983 & 47 & 24.25 & 1419 & 45.51 & 33.80 & 48555 & 40.02 & 70.83 \\
B & 18393 & 53 & 16.44 & 1268 & 18.27 & 10.98 & 43985 & 15.92 & 11.71 \\
C & 23451 & 39 & 3.70 & 2146 & 41.57 & 34.65 & 48864 & 36.17 & 14.65 \\
D & 14305 & 23 & 1.29 & 1528 & 46.39 & 20.57 & 16315 & 44.62 & 2.82 \\
\bottomrule
\end{tabular}

\end{adjustbox}
\end{table}

\clearpage
\bibliographystylesup{unsrtnat}
\bibliographysup{impact}

\end{document}